\documentclass[twocolumn,superscriptaddress,reprint,nofootinbib,prd]{revtex4-1}%
\usepackage{amssymb}
\usepackage{color}
\usepackage{graphicx}
\usepackage{multirow}
\usepackage{dcolumn}
\usepackage{bm}
\usepackage{subfigure}
\usepackage[header,title,page,titletoc]{appendix}
\usepackage{amsmath}
\usepackage{bbm}
\usepackage{amsfonts}%
\usepackage[bookmarks,colorlinks=false]{hyperref}
\usepackage{ulem}
\newlength{\wordlength}

\newlength{\onewordlength}

    \newcommand{\ba}{\begin{eqnarray}}
    \newcommand{\ea}{\end{eqnarray}}
    \newcommand{\be}{\begin{equation}}
    \newcommand{\ee}{\end{equation}}

\newcommand{\bzero}{{\bf 0}}
\newcommand {\btheta} {{\mbox{\boldmath$\theta$}}}

\newcommand{\bfe}{{\bf e}}
\newcommand{\bn}{{\bf n}}

\newcommand {\bq} {{\mathbf q}}

\newcommand {\bp} {{\mathbf p}}
\newcommand{\bx}{{\bf x}}
\newcommand{\by}{{\bf y}}

\newcommand{\calO}{{\mathcal O}}

\newcommand{\beq}{\begin{eqnarray}}
\newcommand{\eeq}{\end{eqnarray}}
\newcommand{\bw}{\begin{widetext}}
\newcommand{\ew}{\end{widetext}}

\begin{document}

\title{Two Photon Decays of $\eta_c$ from Lattice QCD}

\author{Ting Chen}
\affiliation{%
School of Physics, Peking University, Beijing 100871, China
}%

\author{Ying Chen}
\affiliation{%
Institute of High Energy Physics, Chinese Academy of Sciences, Beijing 100049, China
}%

\author{Ming Gong}
\affiliation{%
Institute of High Energy Physics, Chinese Academy of Sciences, Beijing 100049, China
}
\author{Yu-Hong Lei}
\affiliation{%
School of Physics, Peking University, Beijing 100871, China
}%

\author{Ning Li}
\affiliation{%
School of Science, Xi'an Technological
University, Xi'an  710032, China
}%


\author{Chuan Liu}%
\email[Corresponding author. Email: ]{liuchuan@pku.edu.cn}
\affiliation{%
School of Physics and Center for High Energy Physics, Peking
University, Beijing 100871, China
}%
\affiliation{Collaborative Innovation Center of Quantum Matter, Beijing 100871, China}
%
%
%
%
\author{Yu-Bin Liu}
\affiliation{%
School of Physics, Nankai University, Tianjin 300071, China
}
\author{Zhaofeng Liu}
\affiliation{%
Institute of High Energy Physics, Chinese Academy of Sciences, Beijing 100049, China
}
\author{Jian-Ping Ma}
\affiliation{%
Institute of Theoretical Physics, Chinese Academy of Sciences, Beijing 100190, China
}
\author{Wei-Feng Qiu}
\affiliation{%
Institute of High Energy Physics, Chinese Academy of Sciences, Beijing 100049, China
}
\author{Zhan-Lin Wang}
\affiliation{%
School of Physics, Peking University, Beijing 100871, China
}
%
%
\author{Jian-Bo~Zhang}
\affiliation{%
Department of Physics, Zhejiang University, Hangzhou 311027, China
}
 \collaboration{CLQCD Collaboration}
 \begin{abstract}

 We present an exploratory lattice study for the two-photon decay of $\eta_c$ using $N_f=2$ twisted mass lattice QCD gauge configurations generated by the European Twisted Mass Collaboration. Two different lattice spacings of $a=0.067$fm and $a=0.085$fm are used in the study, both of which are of physical size of 2$fm$.
 The decay widths are found to be $1.025(5)$KeV for the coarser lattice and
 $1.062(5)$KeV for the finer lattice respectively where the errors are purely statistical.
 A naive extrapolation towards the continuum limit yields $\Gamma\simeq 1.122(14)$KeV
 which is smaller than the previous quenched result and most of the current experimental results.
 Possible reasons are discussed.
 \end{abstract}
\maketitle

 \section{Introduction}
 Charmonium systems play a major role in the understanding of the foundation of quantum chromodynamics (QCD),
 the fundamental theory for the strong interaction.
 Due to its intermediate energy scale and the special features of QCD,
 both perturbative and non-perturbative physics show up within charmonium physics, making
 it an ideal testing ground for our understanding of QCD from both sides.

 Two-photon decay width of $\eta_c$ has been attracting
 considerable attention over the years from both theory and experiment sides.
 For example, it is related to the process $gg\rightarrow\eta_c$ relevant for charmonia production
 at Large Hadron Collider (LHC) and the
 small-x gluon distribution function from the inclusive production of $\eta_c$ which describes
 the non-leptonic B mesons decays~\cite{Diakonov:2012vb}.
 Furthermore, two-photon branching fraction for charmonium provides a probe for the strong coupling constant at the charmonium scale via the two-photon decay widths, which can be utilized as a sensitive test for the corrections for the non-relativistic approximation in the quark models or
 the effective field theories such as non-relativistic QCD (NRQCD).

 On the experimental side, considerable progress has been made in recent years
 in the physics of charmonia via the investigations from Belle, BaBar, CLEO-c and BES~\cite{Lees:2010de,Pham:2013ida,Savinov:2013hda,Ablikim:2012xi}.
 Two methods can be utilized to measure the two-photon branching fraction for charmonium.
 One is reconstructing the charmonium in light hadrons with two-photon fusion at $e^+ e^-$ machines.
 The other one is to make $p\bar{p}$ pairs annihilated to charmonium with decay and then to detect
 the real $\gamma \gamma$ pairs.
 Improvements of the measurement for two-photon branching fraction of charmonia will soon be reached
 in the future.

 On the theoretical side, charmonium electromagnetic transitions have been investigated using various theoretical methods~\cite{Guo:2014zva,Bondareko:2012qk,Li:2011ssa,Chen:2011kpa,Meissner:2011xj,TheHadronSpectrum:2010apa,Thomas:2010zzf,Kuang:2009zz,Dudek:2009kk}.
 In principle, these processes involve both electromagnetic and
 strong interactions, the former being perturbative in nature while the latter being non-perturbative.
 Therefore, the study for charmonium transitions requires
 non-perturbative theoretical methods such as lattice QCD.
 Normal hadronic matrix element computations are standard in lattice QCD, however, processes
 involving initial or final photons are a bit more subtle. Since photons are not QCD eigenstates,
 one has to rely on perturbative methods to ``replace'' the photon states
 by the corresponding electromagnetic currents that they couple to.
 The details of this idea was illustrated in Ref.~\cite{Ji:2001wha,Ji:2001nf}.
 Using this technique, the first {\it ab initio} quenched lattice calculation of two photon decay of
 charmonia was reported in Ref.~\cite{Dudek:2006ut}. They found a reasonable agreement with the
 experimental world-average values for $\eta_c$ and $\chi_{c0}$ decay rates.
 However, an unquenched lattice study is still lacking. In this paper, we would like to
 fill this gap by exploring the two photon decay rates of $\eta_c$ meson in lattice QCD
 with $N_f=2$ flavors of light quarks in the sea.
 The gauge configurations utilized in this study are generated by the European Twisted
 Mass Collaboration (ETMC)~\cite{Frezzotti:2004wz,Shindler:2007vp,Boucaud:2007uk,Boucaud:2008xu,%
 Blossier:2007vv,Blossier:2009bx,Baron:2010bv,Baron:2009wt,Alexandrou:2008tn,Alexandrou:2009qu,%
 Jansen:2003ir,Blossier:2010cr},
 where the twisted mass fermion parameters are set at the maximal twist.
 This ensures the so-called automatic $\mathcal{O}(a)$ improvement for on-shell observables
 where $a$ is the lattice spacing~\cite{Frezzotti:2001ea}.

 This paper is organized as follows. In Section~\ref{sec:method}, we briefly
 review the calculation strategies for the matrix element for two-photon decay of $\eta_c$.
 The matrix element is normally parameterized using a form factor, which in turn is directly
 related to the double photon decay rates.
 The following section~\ref{sec:simulation_details} are divided into three parts containing details of
  the simulation: In section~\ref{subsec:configurations}, we introduce the twist mass fermion formulation and give the parameters of the lattices used in our simulation. In section~\ref{subsec:masses}, the continuum and lattice dispersion relations for $\eta_c$ are checked. In section~\ref{sec:formfactor}, numerical results of the form factor are presented which are then converted to the decay width of $\eta_c$ meson.
  Our final number comes out to be smaller than the world-average experimental result
  and barely agrees with the previous quenched result. Possible reasons are discussed
  for this discrepancy.
  In Section~\ref{sec:conclude}, we discuss possible extensions of this calculation
  in the future and conclude.

 \section{Strategies for the computation}
 \label{sec:method}

 In this section, we briefly recapitulate the methods for the calculation of two-photon
 decay rate of $\eta_c$ presented in Ref.~\cite{Dudek:2006ut}.
 The amplitude for two-photon decay of $\eta_c$ can be expressed
 in terms of a photon two-point function in Minkowski space by means of
 the Lehmann-Symanzik-Zimmermann (LSZ) reduction formula,
\bw
\begin{multline}
\label{eq:lsz}
 \langle \gamma(q_1, \lambda_1)\gamma(q_2, \lambda_2) | \eta_c(p) \rangle
 = - \lim_{\substack{q'_1 \to q_1 \\ q'_2 \to q_2}} \epsilon^*_\mu(q_1,
 \lambda_1) \epsilon^*_\nu(q_2, \lambda_2)
 q_1'^2 q_2'^2 \int d^4x d^4y \, e^{i q'_1.y +
 i q'_2.x} \langle \Omega | T\big\{ A^\mu(y) A^\nu(x) \big\} |\eta_c(p_f) \rangle.
\end{multline}
\ew
Here $|\Omega\rangle$ designates the QCD vacuum state, $|\eta_c(p)\rangle$ is the
state with an $\eta_c$ meson of four-momentum $p$ and $|\gamma(q_i,\lambda_i)\rangle$ for $i=1,2$
denotes a single photon state with corresponding polarization vector $\epsilon(q_i,\lambda_i)$,
with $q_i$ and $\lambda_i$ being the corresponding four-momentum and helicity, respectively.
Then one utilizes the perturbative nature of the photon-quark coupling to
approximately integrate out the photon fields and rewrites the
corresponding path-integral as,
\bw
\beq
\int {\cal D}A {\cal D} \bar{\psi} {\cal D}\psi e^{i S_{QED}[A, \bar{\psi}, \psi]} A^\mu(y) A^\nu(x)& =&
 \int {\cal D}A {\cal D} \bar{\psi} {\cal D}\psi  e^{i S_0[A, \bar{\psi}, \psi]}
 \big( \ldots + \tfrac{e^2}{2}\int d^4z d^4w \;\nonumber\\
 & &\times\left[\bar{\psi}(z)\gamma^\rho \psi(z) A_\rho(z)\right] \; \left[\bar{\psi}(w)\gamma^\sigma \psi(w) A_\sigma(w)\right]  +\ldots \big)   A^\mu(y) A^\nu(x).
\eeq
\ew
 The integration over the photon fields can be carried out by Wick
 contracting the fields into propagators. Neglecting
 the disconnected diagrams, one arrives at the following equation,
\bw
\beq
\label{eq:big1}
\langle \gamma(q_1, \lambda_1)\gamma(q_2, \lambda_2) | \eta_c(p) \rangle
&=&{(-e^2)} \lim_{\substack{q'_1 \to q_1 \\ q'_2 \to q_2}} \epsilon^*_\mu(q_1,
 \lambda_1) \epsilon^*_\nu(q_2, \lambda_2) q_1'^2 q_2'^2
 \int d^4x d^4y d^4w\ d^4z e^{i q'_1.y +
 i q'_2.x}\, D^{\mu \rho}(y ,z) D^{\nu \sigma}(x, w)\nonumber \\
& &\times\langle
 \Omega | T\big\{ j_\rho(z) j_\sigma(w)\big\} | \eta_c(p_f)\rangle.
\eeq
\ew
In this equation,
\be
D^{\mu\nu}(y,z) = -i
g^{\mu\nu} \int \tfrac{d^4k}{(2\pi)^4}\tfrac{e^{-i k.(y-z)}}{k^2 + i\epsilon}\;,
\ee
 is the free photon propagator,
 which in momentum space will cancel out the inverse propagators outside the
 integral in Eq.~(\ref{eq:lsz}) and Eq.~(\ref{eq:big1}) when the limit is taken.
 Effectively, each initial/final photon state in the problem is replaced by
 a corresponding electromagnetic current operator which couples to the photon
 and eventually one needs to compute a three-point function of the form
 $\langle\Omega | T\big\{ j_\rho(z) j_\sigma(w)\big\} | \eta_c(p_f)\rangle$.
 This quantity is non-perturbative in nature and should be computed
 using lattice QCD methods.

 The current operators such as $j_\rho(x)$ appearing in Eq.~(\ref{eq:big1}) are electromagnetic
 current operators due to all flavors of quarks. However, we will only consider the charm quark in
 this preliminary study. Contributions due to other quark flavors, e.g. up, down or strange, only come in via disconnected diagrams which are neglected in this exploratory study.
 Another subtlety in the lattice computation is that, with $c(x)/\bar{c}(x)$ being the bare charm/anti-charm quark field on the lattice, composite operators such as the current $j_\rho(x)=Z_V(g_0^2)\bar{c}(x)\gamma_{\rho}c(x)$
 needs an extra multiplicative renormalization factor $Z_V$ which we infer from Ref.~\cite{Becirevic:2012dc}.
 To be specific, for the two set of lattices used in this study,
 the values of the renormalization factor $Z_V(g_0^2)$ are $0.6103(3)$ and $0.6451(3)$ for the lattice size $24^3\times48$ at $\beta=3.9$ and $32^3\times64$ at $\beta=4.05$, respectively.
 Annihilation diagrams of the charm quark itself are also neglected
 due to OZI-suppression. In fact, in our twisted mass lattice setup,
 we introduce two different charm quark fields with degenerate masses,
 so that this type of diagram is absent, see subsection~\ref{subsec:configurations}.

 The resulting expression~(\ref{eq:big1}) can then be analytically
 continued from Minkowski to Euclidean space.
 This continuation works as long as none of the $q^2_i$ is too time-like.
 To be precise, the continuation is fine as long as the virtualities of
 the two photons $Q^2_i\equiv (-q^2_i)> -M^2_V$ where $M_V$ is
 the mass of the lightest vector meson in QCD~\cite{Ji:2001wha,Dudek:2006ut}.
 For quenched lattice QCD, the lightest vector meson is $J/\psi$. However, for our
 unquenched study, it is safe to take $M_V=m_\rho$, i.e. the mass of the $\rho$ meson.
 Using suitable interpolating operator (denoted by $\calO_{\eta_c}(x)$) to create
 an $\eta_c$ meson from the vacuum and reversing the operator time-ordering for later convenience,
 we finally obtain,
\bw
\beq
\label{master}
\langle \eta_c(p_f) | \gamma(q_1, \lambda_1) \gamma(q_2, \lambda_2)\rangle
&=&
\lim_{t_f-t \to \infty} e^2 \frac{ \epsilon_\mu(q_1,
 \lambda_1) \epsilon_\nu(q_2, \lambda_2)}{\tfrac{Z_{\eta_c}(\bp_f)}{2 E_{\eta_c}(\bp_f)} e^{-E_{\eta_c}(\bp_f)(t_f-t)}} \int dt_i
e^{- \omega_1 |t_i -t|}\nonumber\\
& &
\left\langle \Omega \left| T\Big\{ \int d^3\bx\, e^{-i\bp_f\cdot\bx}
\calO_{\eta_c}(\bx, t_f) \int d^3 \by\, e^{i\bq_2\cdot\by}
j^\nu(\by, t) j^\mu(\bzero, t_i)\Big\}\right|\Omega \right\rangle,
\eeq
\ew
 where $\calO_{\eta_c}(x)$ is an interpolating operator that will create an
 $\eta_c$ meson from the vacuum and $\omega_1$ is the energy of the first photon.
 The kinematics in this equation is such that four-momentum conservation $p_f=q_1+q_2$ is valid.
 This equation serves as the starting point for our subsequent lattice computation.
 Basically, the current that couples to the first photon is placed
 at the source time-slice $t_i$, the second current is at $t$ while the final $\eta_c$ meson
 is at the sink time-slice $t_f$ and we are led to the computation
 of a three-point function of the form
 $\langle\Omega|\calO_{\eta_c}(\bx,t_f)j^\nu(\by,t)j^\mu(\bzero,t_i)|\Omega\rangle$.
 Of course, one has to compute the above three-point functions for each $t_i$ and
 perform an integration (summation) over $t_i$.

  Apart from the above mentioned three-point functions, we also need information
 from $\eta_c$ two-point function. For example,
 in the above equation, $Z_{\eta_c}(p_f)$ is the spectral weight factor while $E_{\eta_c}(p_f)$ is the
 energy for $\eta_c$ with four-momentum $p_f=(E_{\eta_c},\bp_f)$.
 These can be inferred from the corresponding two-point functions for $\eta_c$.
 For this purpose, two-point correlation functions
 for the interpolating operators $\calO_{\eta_c}$
 are computed in the simulation:
 \begin{widetext}
 \be \label{eq:twopoint}
 C(\bp_f; t)\equiv\sum_{\bx}e^{-i\bp_f\cdot\bx}\langle \Omega| \mathcal
 {O}_{\eta_c}(\bx,t)\mathcal{O}^\dagger_{\eta_c}({\mathbf 0},0)|\Omega\rangle
 \stackrel{t\gg 1}{\longrightarrow}\frac{|Z_{\eta_c}(\bp_f)|^2}{E_{\eta_c}(\bp_f)}e^{-E_{\eta_c}(\bp_f)\cdot\frac
 {T}{2}}\cosh\left[E_{\eta_c}(\bp_f)\cdot\left(\frac T2-t\right)\right]\;,
 \ee
 \end{widetext}
 where $Z_{\eta_c}(\bp_f)=\langle\Omega|\mathcal{O}_{\eta_c}|\eta_c(\bp_f)\rangle$
 is the corresponding overlap matrix element.

\begin{figure*}[!htbp]
\begin{minipage}{0.45\linewidth}
  \centerline{\includegraphics[width=9.0cm]{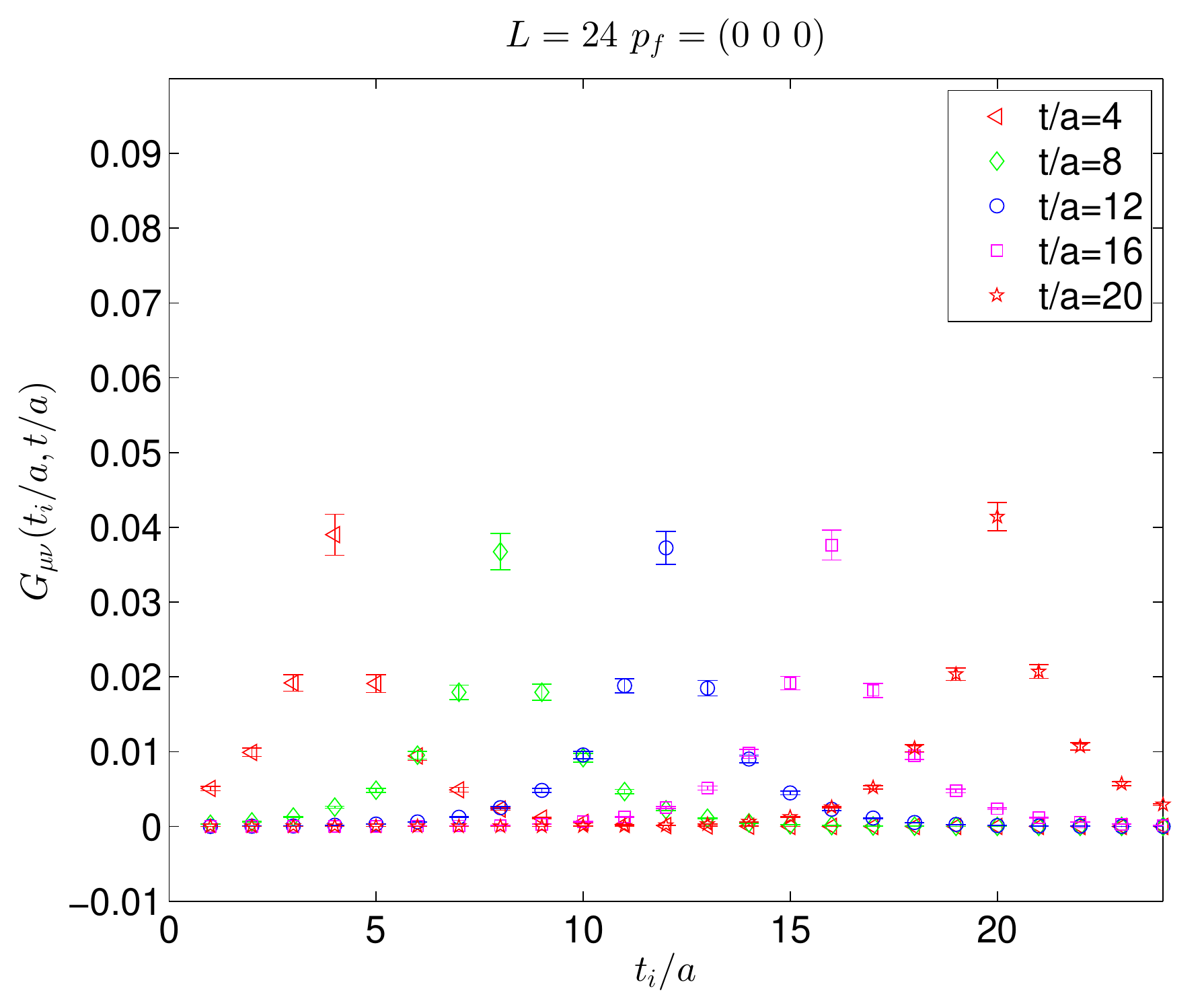}}
  \centerline{$n_2=(0 -1 -2)$; $n_f=(0\ 0\ 0)$}
  \center\ \ \ \ \ \ \ \ \ \ \ \ \ \ \
  \center lattice size: $24^3\times{48}$
\end{minipage}
\hfill
\begin{minipage}{0.45\linewidth}
  \centerline{\includegraphics[width=8.8cm]{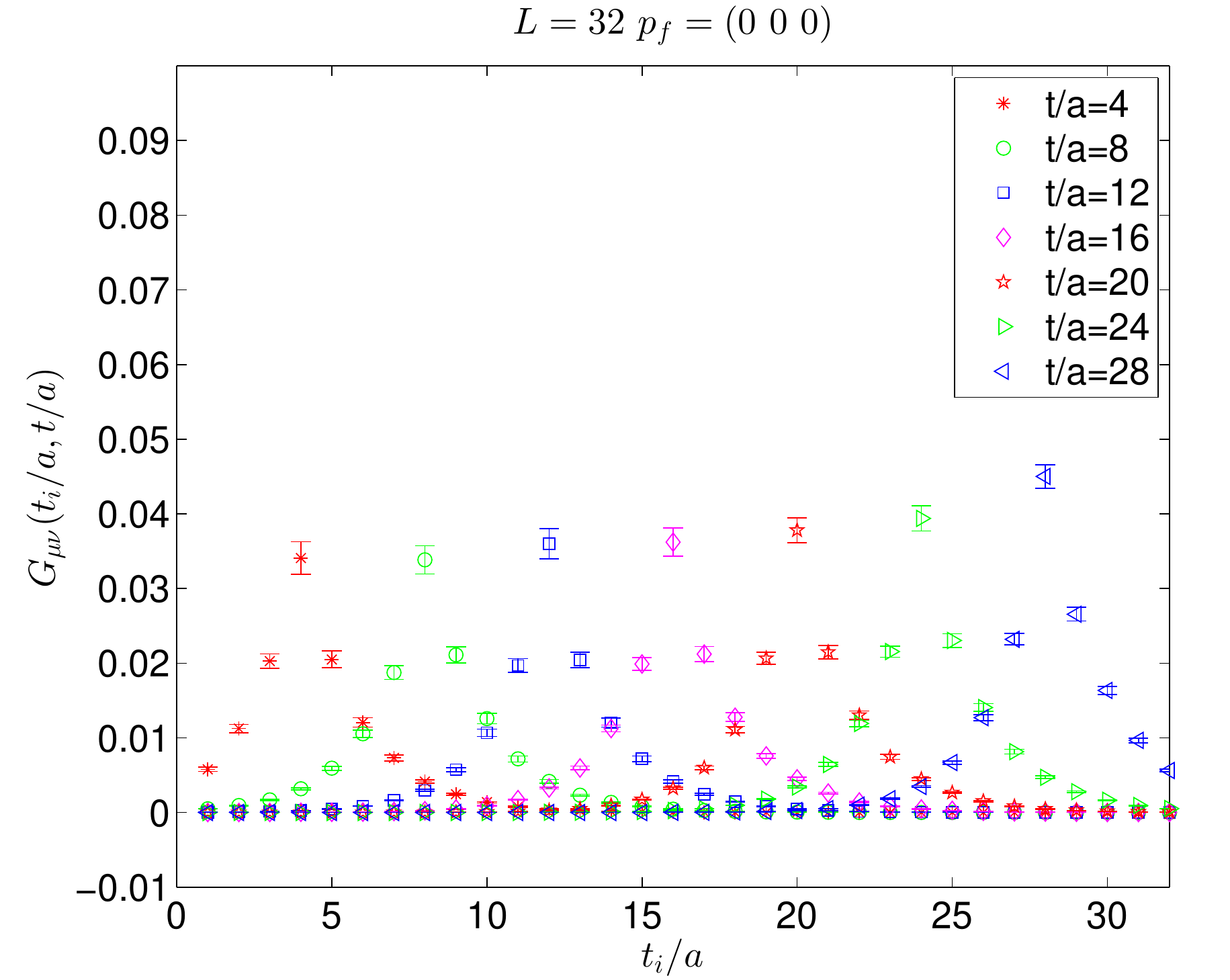}}
  \centerline{$n_2=(0 -1 -2)$; $n_f=(0\ 0\ 0)$}
  \center\ \ \ \ \ \ \ \ \ \ \ \ \ \ \
  \center lattice size: $32^3\times{64}$
\end{minipage}
\caption{The integrand in Eq.~(\ref{master}) versus $t_i$ for various insertion points $t$ obtained from
 our simulation with ensemble I (left panel), and ensemble II (right panel).
 We take $\bn_2=(0,-1,-2)$; $\bn_f=(0,0,0)$ in this example. The insertion points are $t=4,~8,~12,~16,~20$ and $t=4,~8,~12,~16,~20,~24,~28$ for ensemble I and II, respectively.
 }
\label{fig:FormFactor_fit_choose_t}
\end{figure*}

 The three-point functions, denoted by $G_{\mu\nu}(t_i,t)$, that need to be computed in our simulation
 are of the form,
\bw
 \beq
 \label{eq:Gmunu}
 G_{\mu\nu}(t_i,t)
 &=&\left\langle \Omega \left| T\Big\{ \int d^3 \bx e^{-i\bp_f\cdot\bx}
\calO_{\eta_c}(\bx, t_f)\int d^3 \by e^{i\bq_2\cdot\by}
j^\nu(\by, t) j^\mu(\bzero, t_i)\Big\}\right|\Omega\right\rangle.
 \eeq
 \ew
 Keeping the sink of $\eta_c$  fixed at $t_f=T/2$,
 we compute $G_{\mu\nu}(t_i,t)$ across the temporal direction for all $t_i$ and $t$ on our lattices.
 For a fixed $t_i$, one has to use sequential source technique to obtain the $t$ dependence of the
 three-point function. Then, the same calculation is repeated with a varying $t_i$.
 Then, according to Eq.~(\ref{master}), the desired matrix element is obtained by using
 the results of $G_{\mu\nu}(t_i,t)$ for different combinations of $t_i$ and $t$ and integrate
 over $t_i$ with an exponential weight $e^{-\omega_1|t_i-t|}$.
 In practice, the integral is replaced by a summation over $t_i$.
 To explore the validity of this replacement, we have checked the behavior of the
 integrand some of which are illustrated in Fig.~\ref{fig:FormFactor_fit_choose_t}.
 It is seen that these integrand as a function of $t_i$ indeed peak around
 the corresponding $t$ values.

  The matrix element
 $\langle \eta_c| \gamma(q_1,\lambda_1)\gamma(q_2, \lambda_2)\rangle$
 can be parameterized using the form factor $F(Q_1^2, Q_2^2)$ as follows,
 \ba
 \label{eq:formfactor_parametrization}
 \langle \eta_c| \gamma(q_1,\lambda_1)\gamma(q_2, \lambda_2)\rangle
 &=& 2 (\tfrac{2}{3} e)^2 m_{\eta_c}^{-1} F(Q_1^2,Q_2^2) \epsilon_{\mu\nu\rho\sigma}
 \nonumber \\
 &\times&  \epsilon^\mu(q_1,\lambda_1)
 \epsilon^\nu(q_2,\lambda_2) q_1^\rho q_2^\sigma,
 \ea
 where $\epsilon^\mu(q_1,\lambda_1), \epsilon^\nu(q_2,\lambda_2)$ are the
 polarization vectors of the photons while $q_1$ and $q_2$ are the corresponding four-momenta.
 The physical on-shell decay width $\Gamma$ for $\eta_c$ to two photons
 is related to the form factor at $Q^2_1=Q^2_2=0$, which will be referred to as the
 {\em physical point} in the following, via
 \beq
 \Gamma & =& \pi \alpha_{em}^2 \left(\frac{16}{81}\right)
 m_{\eta_c} |F(0,0)|^2,
 \label{decay_width}
 \eeq
 where $\alpha_{em}\simeq (1/137)$ is the fine structure constant.
 Therefore, to extract the physical decay width, we simply compute the corresponding
 three-point functions in Eq.~(\ref{master}) and then extract the
 form factors $F(Q^2_1,Q^2_1)$ at various virtualities close to the physical point.
 Then, we can extract the information for $F(0,0)$ yielding the physical decay width.
 Although the physical decay width is only related to $F(0,0)$, the behavior
 of $F(Q^2_1,q^2_2)$ at non-zero virtualities are also of physical relevance
 when studying processes involving one or two virtual photons.

\section{Simulation details}
\label{sec:simulation_details}

\subsection{Simulation setup}
\label{subsec:configurations}
 In this study, we use twisted mass fermions at the maximal twist.
 The most important advantage of this setup is the so-called
 automatic $\mathcal O(a)$ improvement for the physical quantities.
 To be specific, we use $N_f=2$ (degenerate $u$ and $d$ quark) twisted mass
  gauge field configurations generated by the European Twisted Mass Collaboration (ETMC).
  The other quark flavors, namely strange and charm quarks, are quenched.
  These quenched flavors are introduced as valence quarks using
  the Osterwalder-Seiler (OS) type action~\cite{Osterwalder:1977pc,Frezzotti:2004wz}.
  Following the Refs.~\cite{Frezzotti:2004wz,Blossier:2007vv,Blossier:2009bx},
  in the valence sector we introduce
 three twisted doublets, $(u,d)$, $(s,s')$ and $(c,c')$ with masses
 $\mu_l$, $\mu_s$ and $\mu_c$, respectively. Within each doublet,
 the two valence quarks are regularized in the physical basis
 with Wilson parameters of opposite signs ($r=-r'=1$).
 The fermion action for the valence sector reads
 \begin{eqnarray}
 S&=&(\bar{\chi}_u,\bar{\chi}_d)\left(D_W+m_{crit}+i\mu_l\gamma_5\tau_3\right)\left({\chi_u \atop \chi_d}\right)\nonumber\\
 & +&(\bar{\chi}_s,\bar{\chi}_{s'})\left(D_W+m_{crit}+i\mu_s\gamma_5\tau_3\right)\left({\chi_s \atop \chi_{s'}}\right)\nonumber\\
 & +&(\bar{\chi}_c,\bar{\chi}_{c'})\left(D_W+m_{crit}+i\mu_c\gamma_5\tau_3\right)\left({\chi_c \atop \chi_{c'}}\right).
 \end{eqnarray}
 One can perform a chiral twist to transform the quark fields
 in physical basis to the so-called twisted basis as follows:
 \begin{eqnarray}
 \left({u \atop d}\right )&=&\exp\left(i\omega\gamma_5\tau_3/2\right)\left({\chi_u \atop \chi_d}\right)\nonumber\\
 \left({s \atop s'}\right )&=&\exp\left(i\omega\gamma_5\tau_3/2\right)\left({\chi_s \atop \chi_{s'}}\right)\nonumber\\
 \left({c \atop c'}\right )&=&\exp\left(i\omega\gamma_5\tau_3/2\right)\left({\chi_c \atop \chi_{c'}}\right)
 \label{chiral}
 \end{eqnarray}
 where $\omega=\pi/2$ implements the full twist.

 Two sets of gauge field ensembles are utilized in this work,
 each containing 200 gauge field configurations.
 We shall call them Ensemble I and II respectively. The explicit parameters
 are listed in Table~\ref{configurations}.
 The corresponding renormalization factor $Z_V(g^2_0)$  and the
 valence charm quark mass parameter $\mu_c$ are taken
 from Ref.~\cite{Becirevic:2012dc}.

 \begin{table}[h]
 \centering \caption{Parameters for the gauge ensembles used in this work.
 See Ref.~\cite{Becirevic:2012dc} and references therein for notations.}
 \label{configurations}
 \begin{tabular}{|c|c|c|c|c|c|c|c|}
 \hline
      Ensemble    & $\beta$       &$a$[fm]      &$V/{a^4}$           &$a\mu_{\rm sea}$   &$m_{\pi}$[MeV]
      &$a\mu_c$ & $Z_V(g^2_0)$\\
\hline
 I & $3.9$         &0.085      &$24^3\times{48}$     &0.004           &315              &0.215 & 0.6103(3)\\
\hline
II & $4.05$        &0.067      &$32^3\times{64}$     &0.003           &300              &0.185 & 0.6451(3)\\
\hline
 \end{tabular}
 \end{table}

 For the meson operators, in the physical basis, we use simple quark bi-linears
 such as $\bar{q}\Gamma q$ and the corresponding form in twisted basis
 will be denoted as $\bar{\chi}_q\Gamma'\chi_q$ which can  be readily obtained
 from Eq.~(\ref{chiral}). For later convenience, these are tabulated in table~\ref{tab:operator11} together
 with the possible $J^{PC}$ quantum numbers in the continuum
 and the names of the corresponding particle in the light and the charm sector.
 The current operators that appear in Eq.~(\ref{eq:Gmunu}) are also listed.
 \begin{table}[!htbp]
 \centering
 \caption{Local interpolating operators for vector and pseudo-scalar states
 and the current operators that appear in Eq.~(\ref{eq:Gmunu}) in both physical and twisted basis,
 $\bar{q}\Gamma q=\bar{\chi}_q\Gamma'\chi_q$. The names of the corresponding particle and their
 $J^{PC}$ quantum numbers in the continuum are also listed.
 The index for $i$, $\mu$ and $\nu$ are $1,2,3$.}
 \label{tab:operator11}
 \begin{tabular*}{8cm}{@{\extracolsep{\fill}}|c|c|c|c|c|}
 \hline
                   &$\rho$/$J/\psi$         &$\pi$/$\eta_c$         &$j^\mu$            &$j^\nu$ \\[1mm]\hline
 $\Gamma$          &$\gamma_i$     &$\gamma_5$      &$\gamma_{\mu}$    &$\gamma_{\nu}$  \\[2mm]\hline
 $\Gamma'$         &$\gamma_i$     &$1$             &$\gamma_{\mu}$    &$\gamma_{\nu}$  \\[2mm]\hline
 $J^{PC}$          &$1^{--}$       &$0^{-+}$        &$1^{--}$      &$1^{--}$            \\[2mm]\hline
 \end{tabular*}
 \end{table}

 \subsection{Twisted boundary conditions}
 \label{subsec:TBC}
  In order to increase the resolution in momentum space, particularly close to the physical point
  of $Q^2_1=Q_2^2=0$, it is customary to implement the twisted boundary conditions (TBC)~\cite{Bedaque:2004kc,Sachrajda:2004mi,Ozaki:2012ce,Becirevic:2012dc} in recent
  lattice form factor computations, see e.g.~\cite{Brandt:2011sj}.
  We have also adopted the twisted boundary conditions for the valence quark fields,
  also known as partially twisted boundary conditions.

  The quark field $\psi_\btheta(\bx,t)$, when it is
 transported by an amount of $L$ along
 the spatial direction $i(i=1,2,3)$,
 will change by a phase factor $e^{i\theta_i}$,
 \be
 \label{eq:twistBC}
 \psi_\btheta(\bx+L\bfe_i,t)=e^{i\theta_i}\psi_\btheta(\bx,t)\;,
 \ee
 where $\btheta=(\theta_1,\theta_2,\theta_3)$ is the twisted angle for
 the quark field in spatial directions which can be tuned freely.
 In this calculations, we only twist one of the charm quark field in both vector currents,
 the other charm quark fields remain un-twisted. If we introduce the new quark fields
 \be
 \hat{c}'(\bx,t)=e^{-i\btheta\cdot\bx/L}c'_\btheta(\bx,t)\;,
 \ee
 it is easy to verify that $\hat{c}'(\bx,t)$ satisfy the
 conventional periodic boundary conditions along all spatial
 directions; i.e, $\hat{c}'(\bx+L\bfe_i, t)=\hat{c}'(\bx,t)$ with $i=1,2,3$ if
 the original field $c'_\btheta(\bx,t)$ satisfies the twisted boundary
 conditions~(\ref{eq:twistBC}). For Wilson-type fermions,
 this transformation is equivalent to the replacement of
 the gauge link; i.e,
 \be
 U_{\mu}(x)\Rightarrow \hat{U}_{\mu}(x)=e^{i\theta_\mu a/L} U_{\mu}(x)\;,
 \ee
 for $\mu=0,1,2,3$ and $\theta_\mu=(0,\btheta)$.
 In other words, each spatial gauge link is modified by a $U(1)$-phase.
 Then the current vectors that appear in Eq.~(\ref{eq:Gmunu}) are
 constructed using the hatted and the original charm quark field as,
 \be
 \left\{\begin{aligned}
 j^\nu(\by, t)&=\bar{c}(\by,t)(\gamma_{\nu})\hat{c}'(\by,t),
 \\
 j^\mu(\bzero, t_i)&=\overline{\hat{c}'}(\bzero,t_i)(\gamma_{\mu}){c}(\bzero,t_i).
 \end{aligned}\right.
 \ee
 The allowed momenta on the lattice are thus modified to
 \be
 \bq_i=\left({2\pi\over L}\right)
 \left(\bn_i+{\btheta\over {2\pi}}\right),
 \ee
 for $i=1,2$ where $\bn_i\in \mathbb{Z}^3$ is a three-dimensional integer.
 By choosing different values for $\btheta$, we
 could obtain more values of $\bq_1$ and $\bq_2$
 than conventional periodic boundary conditions.
 In this paper, apart from the untwisted case of $\btheta=(0,0,0)$,
 we have also computed the following cases: $\btheta=(0,0,\pi)$, $(0,0,\pi/2)$,
 $(0,0,\pi/4)$ and $(0,0,\pi/8)$. These choices offer us many more data points
 in the vicinity of the physical kinematic region.

\subsection{Meson spectrum and the dispersion relations}
 \label{subsec:masses}

 \begin{figure*}[htbp]
\begin{minipage}{0.45\linewidth}
  \centerline{\includegraphics[width=9.0cm]{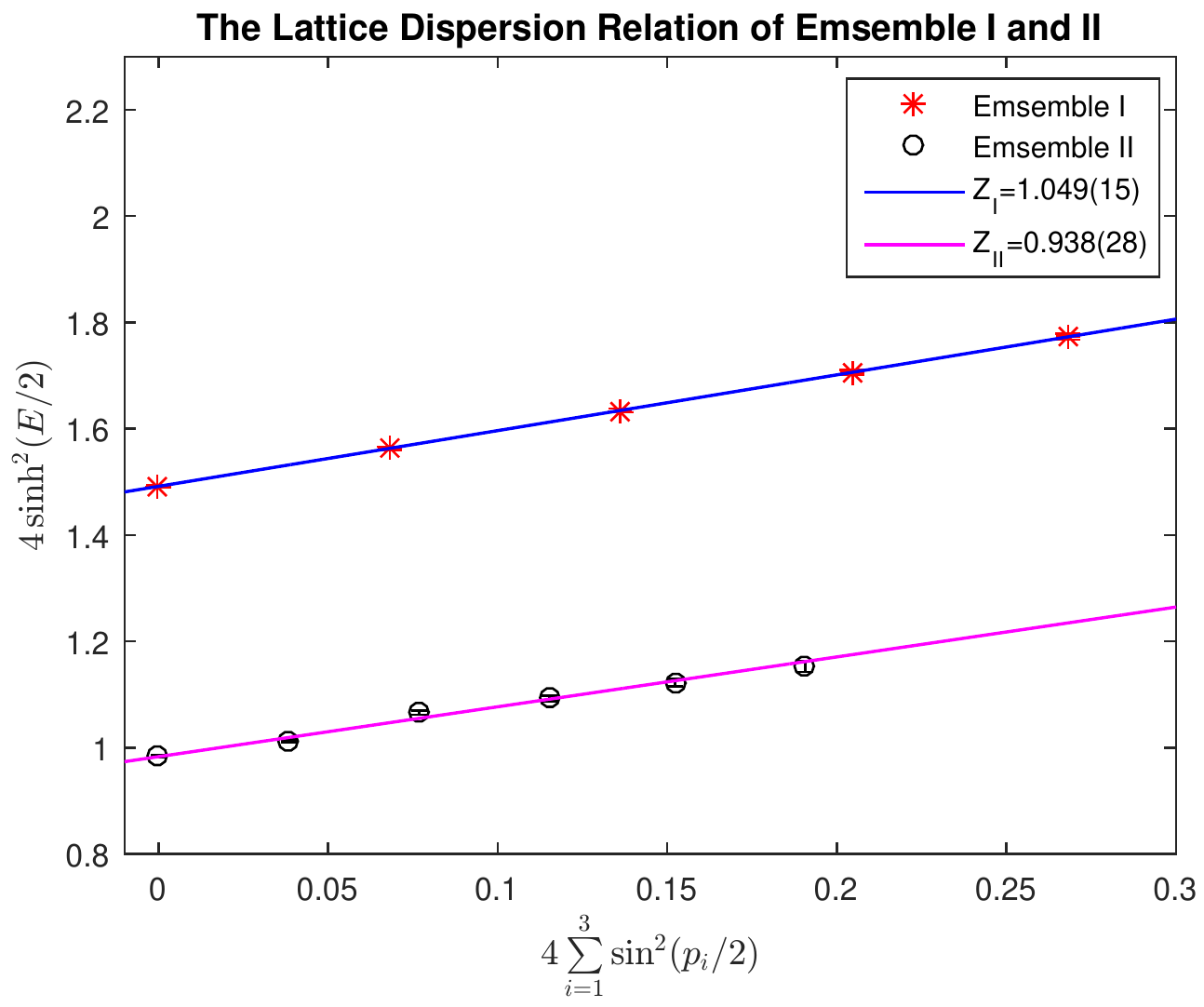}}
\end{minipage}
\hfill
\begin{minipage}{0.45\linewidth}
  \centerline{\includegraphics[width=8.8cm]{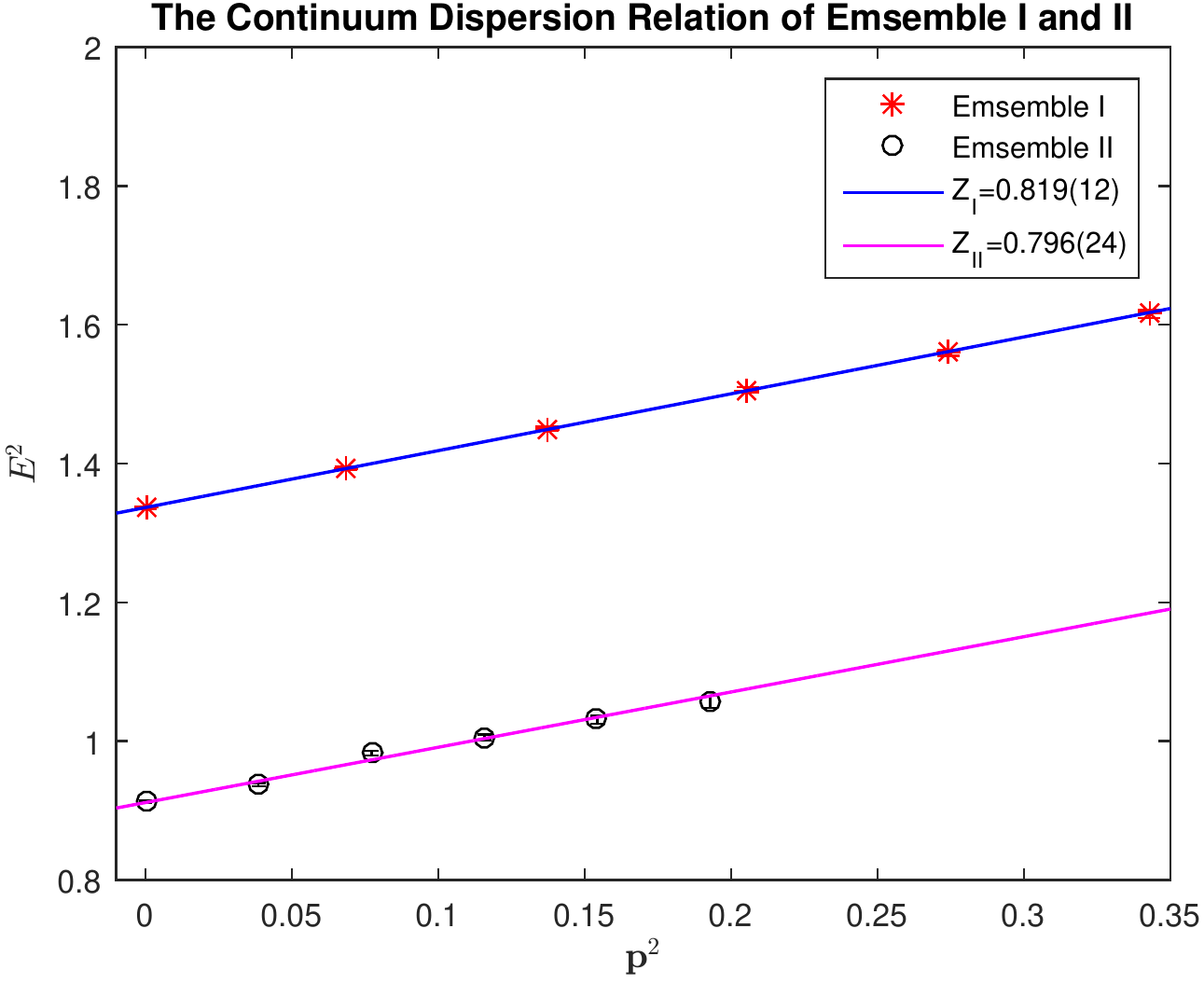}}
\end{minipage}
\caption{$\eta_c$ meson dispersion relation obtained from simulations on Ensemble I (crosses) and
 Ensemble II (open circles).
 In the left/right panel, the horizontal axis represents the lattice/continuum three-momentum squared variable.
 Straight lines in both panels are the corresponding linear fits using lattice/continuum dispersion
 relations Eq.~(\ref{eq:continuum_disper}) or Eq.~(\ref{eq:lattice_disper}).
 The fitted parameters $Z_{I}$ and $Z_{II}$ together with their errors are also shown.}
\label{fig:dispersion_relation}
\end{figure*}
 Before calculating the matrix element with two photon decay from $\eta_c$, the mass for $\eta_c$ and $\rho$ state and
 the energy dispersion relations for $\eta_c$ must be verified. This is particularly important for our study
 due to the following reasons.
 Firstly, we use the $N_f=2$ twisted mass configurations, the sea quarks contains $u$ and $d$ quark field.
 therefore virtual $\rho$ state can enter the game.
 Thus, we should calculate the $\rho$ mass so as to ensure the photon
 virtualities $Q_1^2,Q_2^2>-m_{\rho}^2$ in this simulations.
 Secondly, we do need the information from $\eta_c$ correlation functions, the value of $E_{\eta_c}(\bp)$ and
 $Z_{\eta_c}(\bp)$ in order to extract the relevant matrix elements. Finally, we should also
 check the dispersion relation of $\eta_c$ which is quite heavy in lattice units (around $0.95$) in our simulation
 and therefore some kinematic factors ($q^\rho_1$ and $q^\sigma_2$) that
 enter Eq.(\ref{eq:formfactor_parametrization}) might need modifications accordingly.

 Following Eq.~(\ref{eq:twopoint}), the energy $E_{\eta_c}(\bp_f)$ for $\eta_c$ state
 with three-momentum $\bp_f$ can be obtained from
 the corresponding two-point function via
 \be
 \cosh\left(E_{\eta_c}(\bp_f)\right)=\frac{C(\bp_f;t-1)+C(\bp_f;t+1)}{2C(\bp_f;t)}
 \;.
 \label{two_point}
 \ee
 The two point function is symmetric about $t=T/2$. In real simulation
 we average the data from two halves about $t=T/2$ to improve statistics.
 We use the effective mass plateaus at zero three-momentum for the $\eta_c$ and $\rho$ state
 to obtain the masses which are then listed in Table~\ref{tab:charmmass}.
 The mass of the $\eta_c$ comes out to be lighter than its physical value
 since these values are still finite lattice spacing values. When extrapolated
 towards the continuum limit, the mass will become compatible with the experimental value.
 The mass of the $\rho$ here serves to restrict our kinematic regions
 where analytic continuation is justified.
 \begin{table}[h]
 \centering \caption{The meson mass values
  for $\eta_c$ and $\rho$ obtained from the two ensembles in this work.}
 \label{tab:charmmass}
 \begin{tabular*}{8cm}{@{\extracolsep{\fill}}|c|r|r|}
 \hline
             Ensemble                  & $m_{\eta_c}$[MeV]      & $m_\rho$[MeV]    \\ \hline
 I    &2678(3)       &903(88)      \\ \hline
 II    &2812(2)         &1051(50)          \\ \hline
 \end{tabular*}
 \end{table}

\begin{figure*}[htbp]
\includegraphics[width=0.6\textwidth]{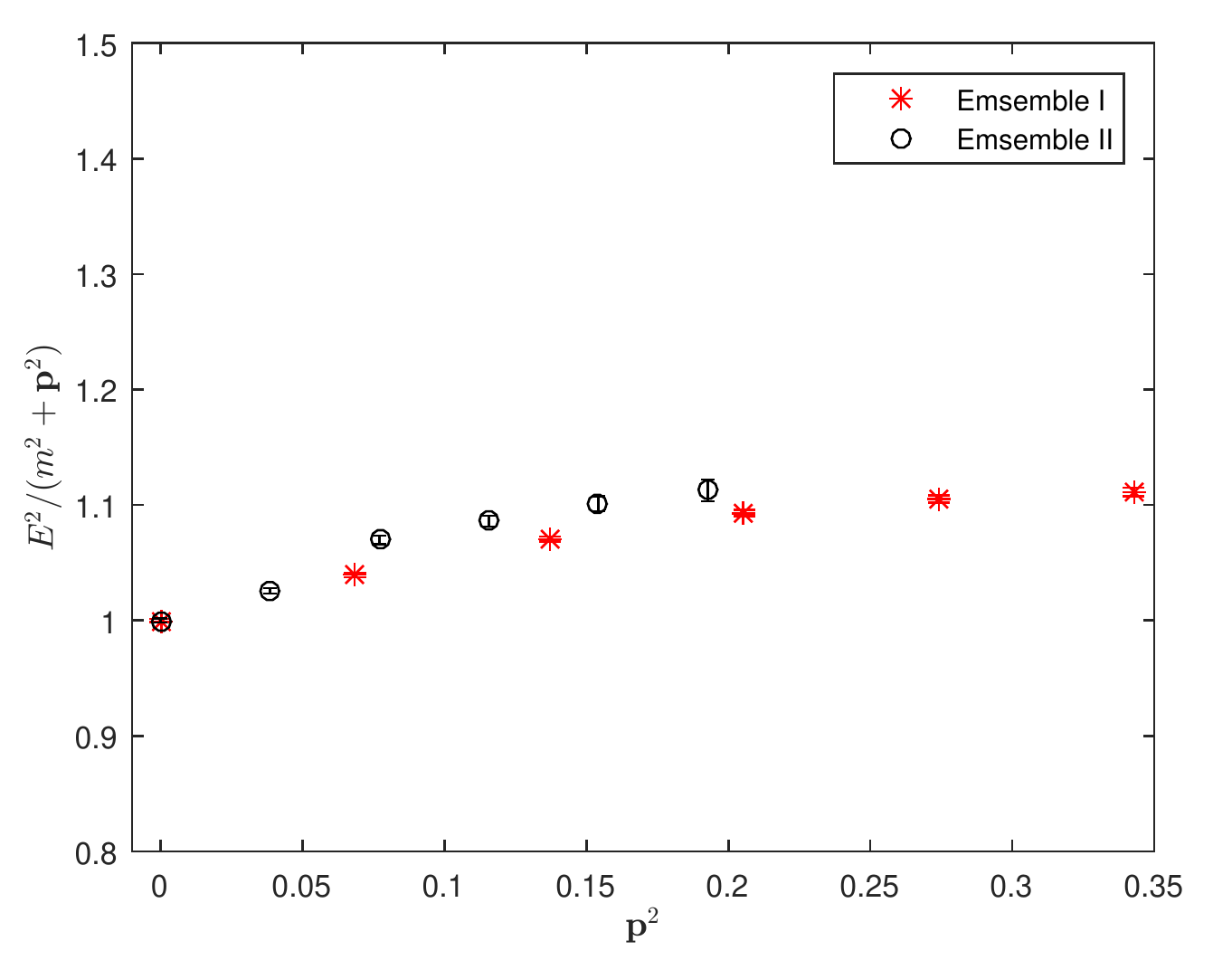}
\caption{The quantity $E^2(\bp)/(m^2+\bp^2)$ is plotted vs. $\bp^2$ in lattice units for the $\eta_c$ meson for
two of our ensembles.
 It is seen that it deviates from unity at rather small values of $\bp^2$ which is caused by
 the difference between the rest mass and kinetic mass of the meson.
\label{fig:dispersion_relation2}}
\end{figure*}

 Similarly, we obtain the energies for $\eta_c$ at non-vanishing momenta
 via Eq.~(\ref{two_point}) which then can be utilized to verify the the following
 two dispersion relations: the conventional one in the continuum,
 \be
 \label{eq:continuum_disper}
 E^2(\bp) =m^2 +Z_{\rm cont}\cdot\sum_i{p_i^2}\;,
 \ee
 and its lattice counterpart,
  \be
 \label{eq:lattice_disper}
 4\sinh^2\frac {E(\bp)}{2} =4\sinh^2 \frac{m}{2}
 +Z_{\rm latt}\cdot 4\sum_i\sin^2\left(\frac {p_i} 2\right)\;.
 \ee
 For free particles, the constants $Z_{\rm cont}$ and $Z_{\rm latt}$ should be close to unity.
 In Fig.~\ref{fig:dispersion_relation}, we show this comparison for the two dispersion relations
 of the $\eta_c$ states in our simulation.
 In the left/right panel, the dispersion relations are illustrated using lattice/continuum dispersion
 relations, respectively.
 In both panels, points with errors  are from simulations
 on $32^3\times 64$ (open circles) or $24^3\times 48$ (stars) lattices.
 Straight lines are the corresponding linear fits to the data.
 It is seen that, although both dispersion relations can be fitted nicely using
 linear fits, the slope for the naive continuum dispersion relation, i.e. $Z_{\rm cont}$ is
 definitely different from unity, see e.g. right panel of Fig.~\ref{fig:dispersion_relation},
 while its lattice counterpart $Z_{\rm latt}$ is close.
 This suggests that, for the $\eta_c$ state, we should use the lattice dispersion relations
 instead of the naive continuum dispersion relation.
 This is not surprising  since $\eta_c$ is quite heavy in lattice units.
 This modification of the dispersion relation does have consequences on our determination
 of the form factor.

 To illustrate this difference further, we plot the quantity
 $E^2(\bp)/(m^2+\bp^2)$ as a function of $\bp^2$ in lattice units (i.e. $a^2\bp^2$, note that
 at these small values of $\bp^2$, the difference
 between $\bp^2$ and the lattice version $\hat{\bp}^2$ is negligible) for two of our ensembles.
 This is shown in Fig.~\ref{fig:dispersion_relation2}. It is seen that this
 quantity deviates from unity by as much as  $10$\% even at rather small values
 of $\bp^2$.  This is actually caused by
 the difference between the rest mass and the kinetic mass of the $\eta_c$ meson.

\subsection{Kinematics}
 \label{subsec:kinematic_scanning}

 In order to fully explore the form factor close to the physical point $Q^2_1=Q^2_2=0$,
 we performed a parameter scan in the two virtualities.
 The following notations will be utilized. First of all, in the continuum,
 we will use $q_{1,2}$ to designate the four-momentum of the two photons.
 We will also use $\omega_{1,2}$ to denote the temporal component of $q_{1,2}$,
 i.e. $\omega_{1,2}\equiv q^0_{1,2}$. When the photons are on-shell, we have
 $\omega_{1,2}=|\bq|_{1,2}$ with $\bq_{1,2}$ being the corresponding three-momentum.
 The so-called virtuality of the photons are defined
 as the corresponding four-momentum squared: $Q^2_{1,2}\equiv (-q^2_{1,2})$.

 On the lattice, however, there are also lattice counterparts of the above notations,
 arising from the lattice dispersion relation~(\ref{eq:lattice_disper}).
 For that we simply add a hat on the corresponding variable.
 For example, we will use $\hat{\omega}_{1}=2\sinh(\omega_1/2)$ to denote
 the lattice version of $\omega_1$.

 The computation has to cover the physical interesting kinematic region.
 For this purpose, we have to scan the corresponding parameter space.
 We basically follow the following strategy: We first fix the four-momentum
 of $\eta_c$, $p_f=(E_{\eta_c},\bp_f)$, and place it on a given
 time-slice $t_f=T$. Note that we just have to fix $\bp_f=\bn_f(2\pi/L)$ and
 $E_{\eta_c}$ can be obtained from the dispersion relation~(\ref{eq:lattice_disper}).
 This effectively puts $\eta_c$ on-shell. Here we also have the freedom to pick a value
 for the twist angle $\btheta$. Then, we judiciously choose several values
 of virtuality $Q^2_1$ around the physical point $Q^2_1=0$.
 To be specific, we picked the range $Q^2_1\in [-0.5,+0.5]$GeV$^2$,
 which satisfies the constraint $Q^2_1>-m^2_\rho$.
 \footnote{This is valid with the physical $\rho$ meson mass. Our lattice
 values yield a less stringent constraint.}
 Since $\bp_f=\bq_1+\bq_2$, this means that, for a given $\bp_f$,
 a choice of $\bq_1$ completely specifies $\bq_2$ and vice versa.
 We therefore take several choices of $\bq_1=\bn_1(2\pi/L)$ by
 changing three-dimensional integer $\bn_1$.
 At this stage, we can compute the energy of the first photon $\omega_1$,
 since $\omega^2_1=\bq^2_1-Q^2_1$. It turns out that we can also
 compute the virtuality of the second photon,
 $Q^2_2=|\bq_2|^2-\omega^2_2$,
 since $\omega_2=E_{\eta_c}-\omega_1$ and $\bq_2$ is
 also known by the choice of $\bq_1$.
 One has to make sure that the values of $Q^2_2$ thus computed
 do satisfy the constraint $Q^2_2>-m^2_\rho$ otherwise it is omitted.
 This procedure is summarized as follows:
 \begin{enumerate}
 \item Pick $\bp_f$ and $\btheta$.
 Obtain $E_{\eta_c}(\bp_f)$ from dispersion relation~(\ref{eq:lattice_disper});
 \item Judiciously choose several values of $Q^2_1$ in a suitable range,
 say $Q^2_1\in[-0.5,+0.5]$GeV$^2$;
 \item Pick values of $\bn_1$ such that $\bq_1=\bn_1(2\pi/L)$.
 This fixes both $\omega_1$ and $Q^2_2$, using energy-momentum conservation;
 \item Make sure all values of $Q^2_1,Q^2_2 >-m^2_\rho$, otherwise the choice
 is simply ignored;
 \item For each validated choice above, compute the three-point functions~(\ref{eq:Gmunu}),
 the two-point functions~(\ref{eq:twopoint}) and eventually obtain the
 hadronic matrix element using Eq.~(\ref{master}).
 \end{enumerate}

\subsection{Form factors}
\label{sec:formfactor}

 In order to compute the desired hadronic  matrix element $\langle\eta_c(p_f)|\gamma(q_1,\lambda_1)\gamma(q_2,\lambda_2)\rangle$
 in Eq.~(\ref{master}),
 we choose to place $\eta_c$ state at a
 fixed sink position $t_f=T/2$.  This sink position is then
 used as a sequential source for a backward charm propagator inversion.
 We compute this with all possible source positions $t_i$ and insertion point $t$.
 This method allows us to freely vary the value of $\omega_1$,  $Q^2_1$ (as discussed in previous subsection)
 and to directly inspect the behavior of the integrand in Eq.~(\ref{master}).

 Taking $\bp_f=\bzero$ for the $\eta_c$ state as an example,,
 we show the behavior of the integrand in Fig.~\ref{fig:FormFactor_fit_choose_t} for
 insertion positions $t=4,8,12,16,20$ for ensemble I and $t=4,8,12,16,20,24,28$ for ensemble II.
 It is seen that the integrand is peaked around $t_i=t$, making the contributions close
 to this point the dominant part of the matrix element.
 For the lattice theory, the integration of $t_i$ in Eq.~(\ref{master}) is replaced by
 a summation.
\begin{figure*}
\begin{minipage}{0.45\linewidth}
  \centerline{\includegraphics[width=9.0cm]{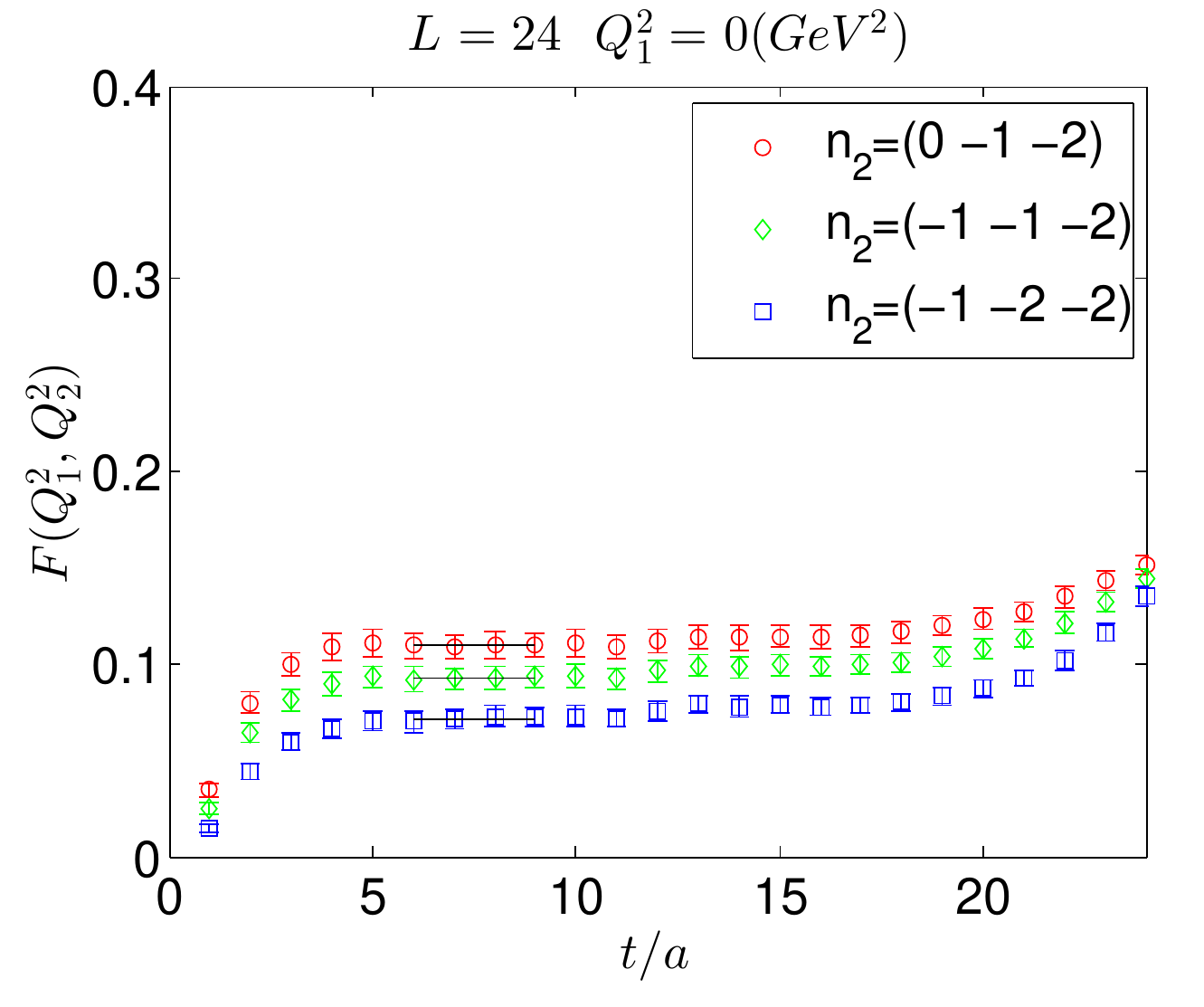}}
  \centerline{Ensemble I}
  \end{minipage}
\hfill
\begin{minipage}{0.45\linewidth}
  \centerline{\includegraphics[width=8.8cm]{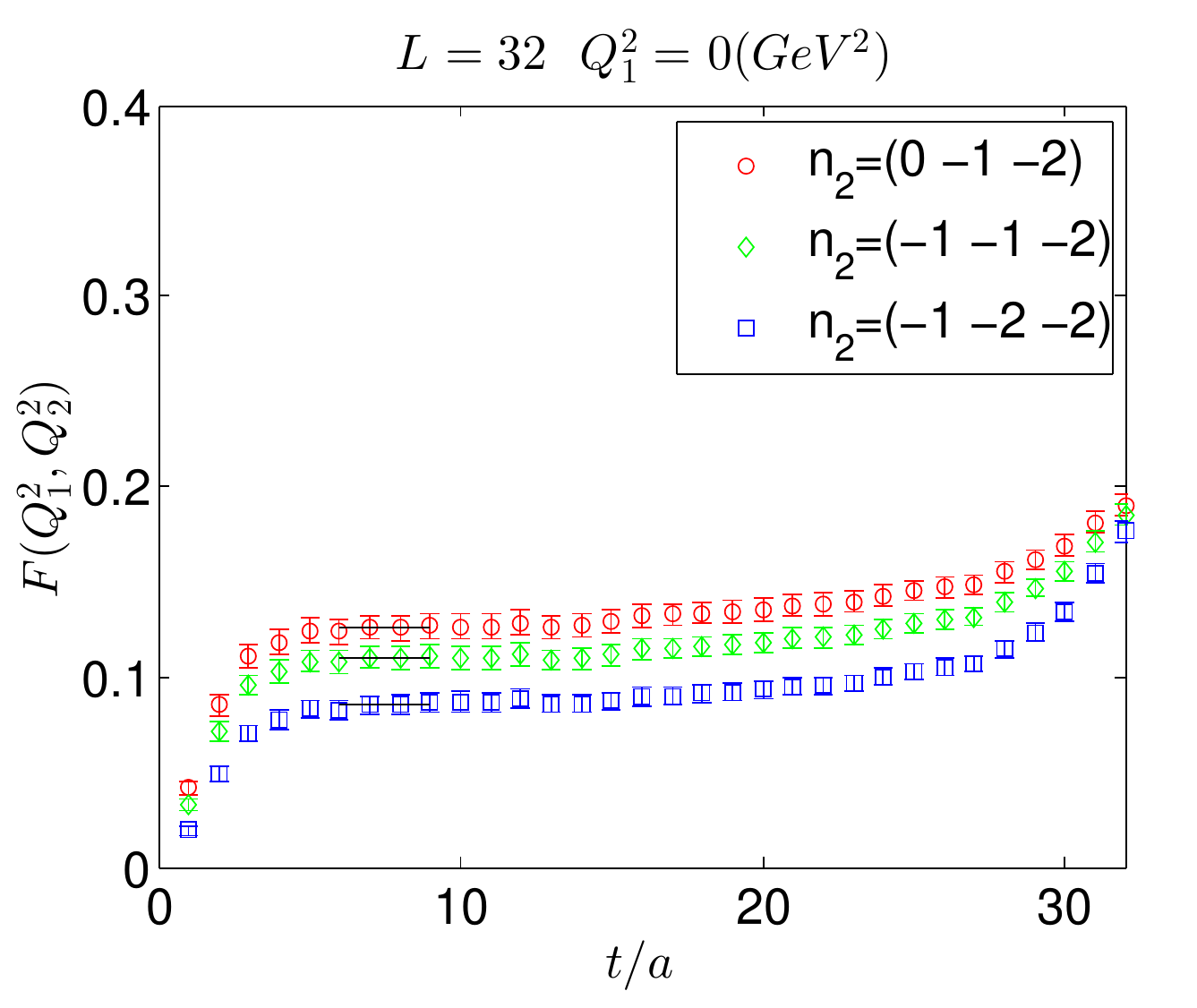}}
  \centerline{Ensemble II}
  \end{minipage}
\caption{The plateau of the form factor obtained by an integration (summation) over $t_i$ for
 three-point function $G_{\mu\nu}(t_i,t)$ with ensemble I (left panel)
 and ensemble II (right panel). We take $Q_1^2=0$; $n_f=(0,0,0)$ in this particular plot.
 Different data points correspond to different choices of $\bn_2$ as indicated.
 }
\label{fig:FormFactor_plateau}
\end{figure*}

 When passing from the matrix element to the form factors,
 one should be careful about the form of the momenta to use. Recall that
 these momentum factors originate from derivatives in the continuum.
 On the lattice, they should be replaced by the corresponding finite differences,
 i.e. one should use the lattice version of the momentum: $q^0\rightarrow 2\sinh(q^0/2)$
 and $q^i\rightarrow 2\sin(q^i/2)$. Since the spatial momenta that we are using
 are relatively small in lattice units, the effect of this replacement might be optional.
 However, for the $0$-th component,
 since each of the photon is roughly half of the $\eta_c$ energy which is large in lattice
 units as we discussed in subsection~\ref{subsec:masses}, this replacement does make
 a difference.

 According to Eq.~(\ref{master}), the matrix element and therefore also the form factor $F(Q^2_1,Q^2_2)$
 should be independent of the insertion point $t$.
 We indeed observe this plateau behavior in our data which is illustrated
 in Fig.~\ref{fig:FormFactor_plateau} for the case of $Q_1^2=0$ as an example.
 Other cases are similar. Fitting these plateaus then yields the corresponding values
 for the matrix element
 $\langle \eta_c| \gamma(q_1,\lambda_1)\gamma(q_2, \lambda_2)\rangle$
 or equivalently the form factor $F(Q^2_1,Q^2_2)$.

 \begin{figure*}[htb]
\begin{minipage}{0.45\linewidth}
  \centerline{\includegraphics[width=9.0cm]{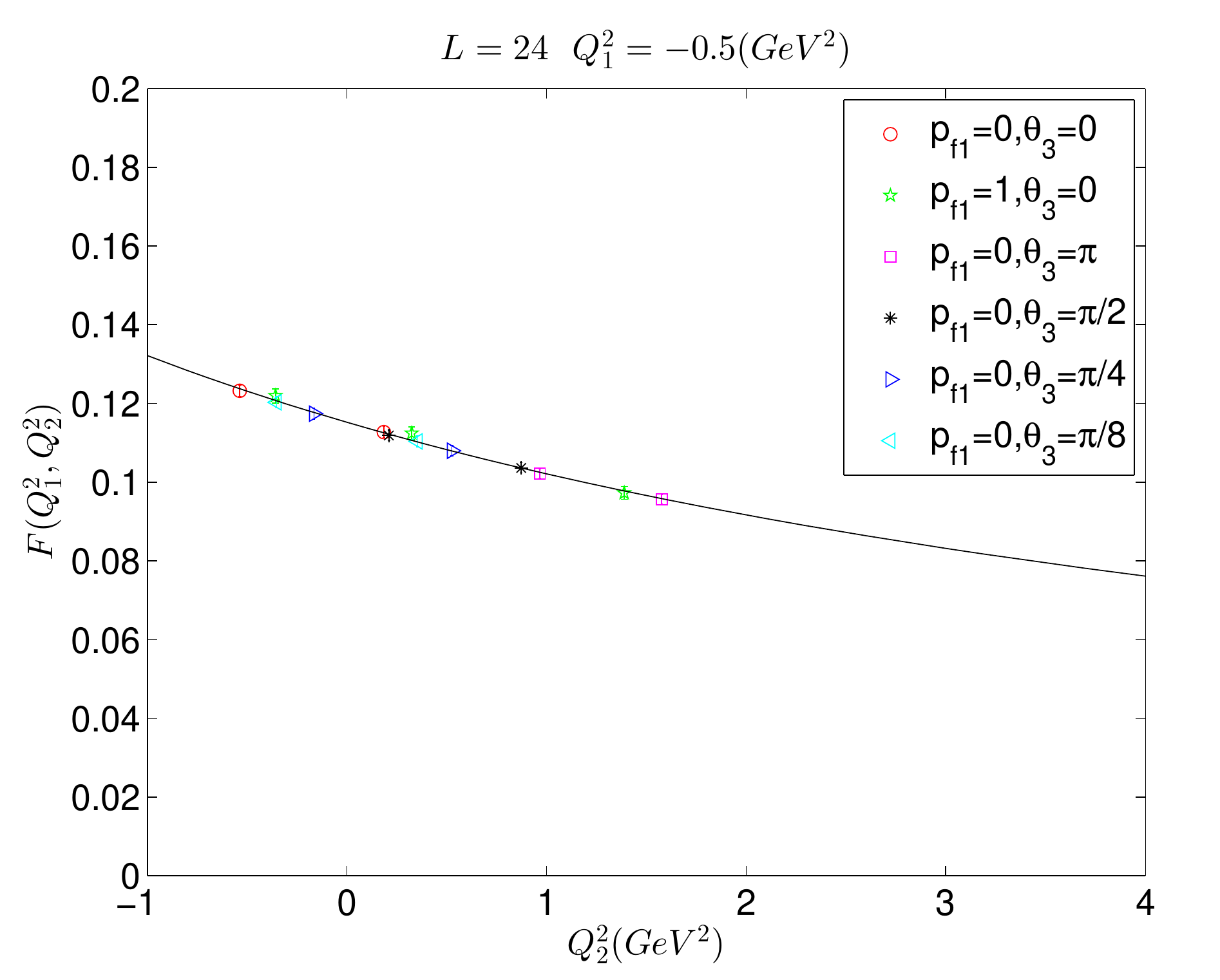}}
  \centerline{Ensemble I}
   \end{minipage}
\hfill
\begin{minipage}{0.45\linewidth}
  \centerline{\includegraphics[width=8.8cm]{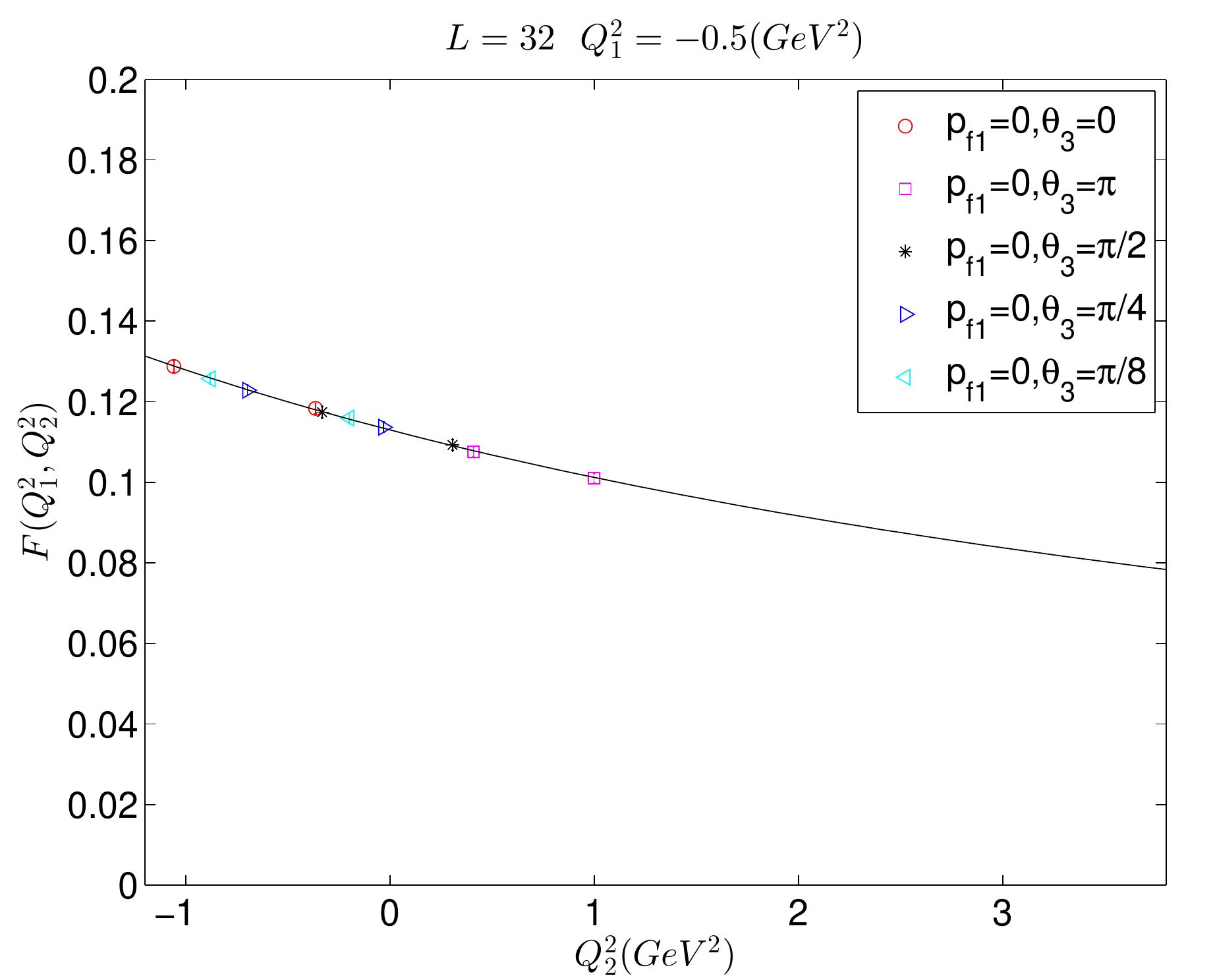}}
  \centerline{Ensemble II}
  \end{minipage}
\caption{The fitted results for $F(Q_1^2, Q_2^2) = F(Q_1^2, 0)/(1+ Q_2^2/\mu^2(Q_1^2))$
 by one-pole form factor for Ensemble I (left figure)
 and Ensemble II (right figure) at a fixed value of $Q^2_1=-0.5$GeV$^2$.
 Different data points correspond to different parameter combinations
 as indicated in the Figure. $p_{f1}$ denotes x-component of the momentum $\bp_f$ of $\eta_c$, and $\theta_3$ represents z-component of twisted angle $\btheta$.}
\label{fig:FormFactor_fit}
\end{figure*}
 To describe the virtuality dependence of the form factor,
 we adopt a simple one-pole parametrization to fit our data.
 \be
 \label{eq:formfactor_fit_1}
 F(Q_1^2, Q^2_2)=F(Q^2_1,0)/(1+Q^2_2/\mu^2(Q^2_1))\;,
 \ee
 where $F(Q^2_1,0)$ and $\mu^2(Q^2_1)$ are regarded as the fitting parameters
 at the given value of $Q^2_1$. Since measurements at different values
 of $Q^2_1$ or $Q^2_2$ are all obtained on the same set of ensembles, we
 adopt the correlated fits, taking into account possible correlations
 among different $Q^2$ values. The covariance matrix among them are
 estimated using a bootstrap method.

 As an example, taking $Q_1^2=-0.5$GeV$^2$, the fitting results are shown
 in Fig.~\ref{fig:FormFactor_fit}. It is seen that this simple formula describes the
 data rather well even for quite large values of $Q^2_2$. We therefore have taken
 all available values of $Q^2_2$ into the fitting process.
 Notice also that, by using the twisted boundary conditions together with different
 combinations of the lattice momenta, we are able to populate the physical region close to
 $Q^2_1=Q^2_2=0$ rather effectively.
 We have tried both correlated and uncorrelated fits on our data.
 The central values for the fitted parameters are compatible, however,
 the error estimates are somewhat different. We adopt the correlated fits
 as our final results.
 Fits for other set of parameters are similar and the final results are summarized
 in Table~\ref{tab:formfactor_fit} for reference.

 \begin{table*}[htb]
\begin{center}
\caption{The summary of fitted results for
 $F(Q_1^2,0)$ and $\mu^2(Q_1^2)$ using Eq.~(\ref{eq:formfactor_fit_1})
 for Ensemble I (left four columns)
 and Ensemble II (right four column).
 The total $\chi^2$ value and the corresponding total degrees of freedom
 is also listed in the columns labelled by $\chi^2/dof$.
 }
  \label{tab:formfactor_fit}
\begin{tabular}{|c|c|c|c|c|c|c|c|}
\hline
\multicolumn{4}{|c|}{Ensemble I}&\multicolumn{4}{|c|}{Ensemble II}\\
\hline\hline
$Q_1^2(GeV^2)$ &$F(Q_1^2,0)$ &$\mu^2(Q_1^2)(GeV^2)$  &$\chi^2/dof$ &$Q_1^2(GeV^2)$ &$F(Q_1^2,0)$ &$\mu^2(Q_1^2)(GeV^2)$ &$\chi^2/dof$ \\
\hline
-0.5      &0.11521(46)    &7.79(38)  &2.22/11   &-0.5        &0.11297(44)  &8.59(42) &0.27/8\\
\hline
-0.4      &0.11353(42)    &7.82(36)  &2.33/11  &-0.4         &0.11163(47)  &8.62(49) &0.24/7\\
\hline
-0.3      &0.11187(39)    &7.83(36)  &2.30/11     &-0.3      &0.11031(49)  &8.61(54) &0.20/6\\
\hline
-0.2      &0.11038(41)    &7.83(37)  &2.12/10     &-0.2      &0.10901(52)  &8.69(53) &0.19/6\\
\hline
-0.1      &0.10874(39)    &7.87(36)  &2.30/10    &-0.1       &0.10771(54)  &8.62(64) &0.15/5\\
\hline
0         &0.10721(43)    &7.90(45)  &1.79/8     &0          &0.10645(59)  &8.67(59) &0.15/5\\
\hline
0.1       &0.10581(46)    &7.82(52)  &1.37/7      &0.1       &0.10523(67)  &8.74(60) &0.13/5\\
\hline
0.2       &0.10432(44)    &7.85(49)  &1.54/7     &0.2        &0.10402(71)  &8.70(73) &0.03/3\\
\hline
0.3       &0.10283(44)    &7.92(47)  &1.43/7    &0.3         &0.10191(56)  &7.74(47) &2.9/3\\
\hline
0.4       &0.10142(45)    &7.96(44)  &1.51/7     &0.4        &0.10056(58)  &7.83(45) &2.8/3\\
\hline
0.5       &0.10012(51)    &7.82(51)  &0.95/5     &0.5        &0.09936(63)  &7.64(55) &2.3/2\\
\hline
\end{tabular}
\end{center}
\end{table*}

\begin{figure*}
\begin{minipage}{0.45\linewidth}
  \centerline{\includegraphics[width=9.0cm]{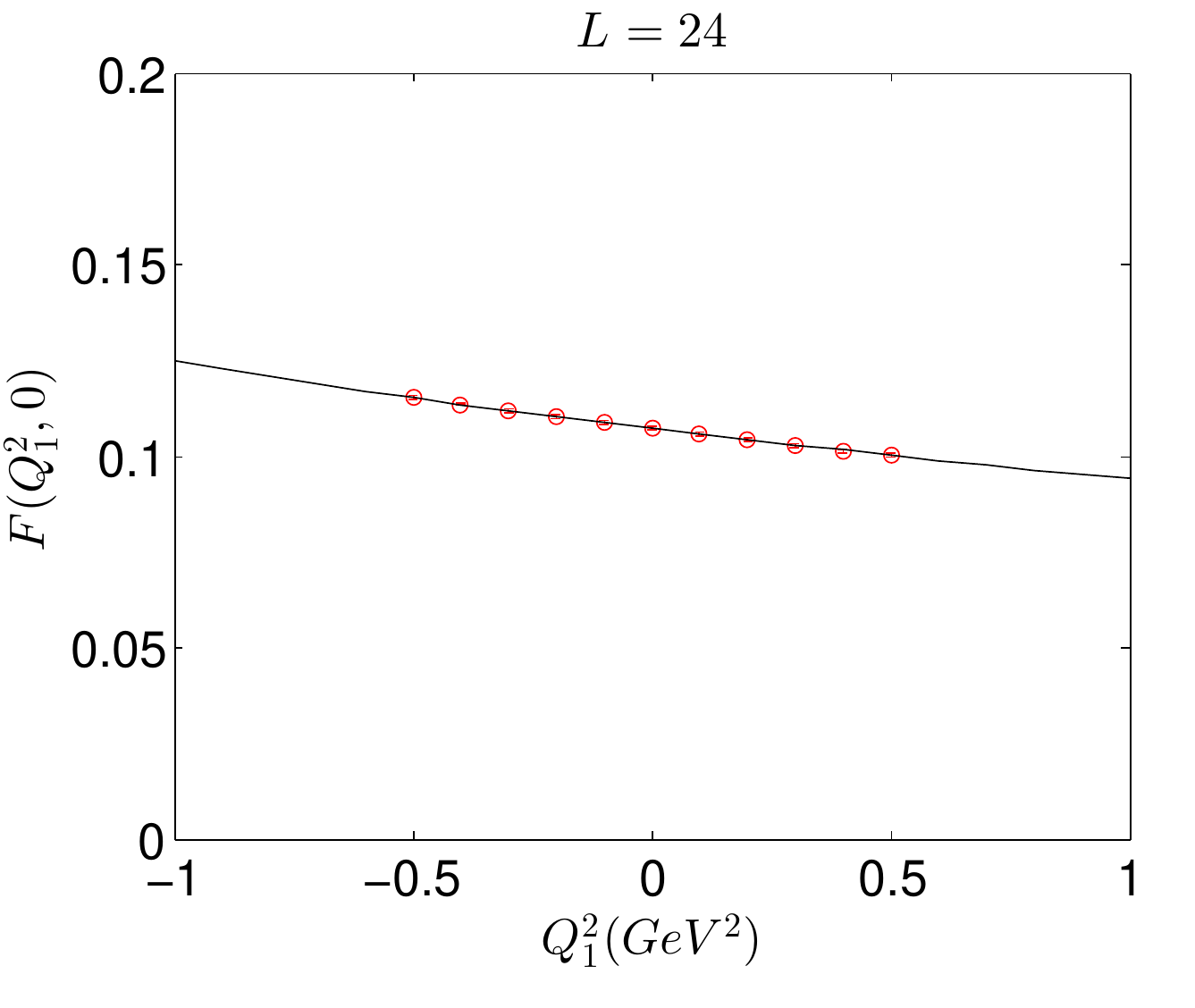}}
  \centerline{Ensemble I}
  \end{minipage}
\hfill
\begin{minipage}{0.45\linewidth}
  \centerline{\includegraphics[width=8.5cm]{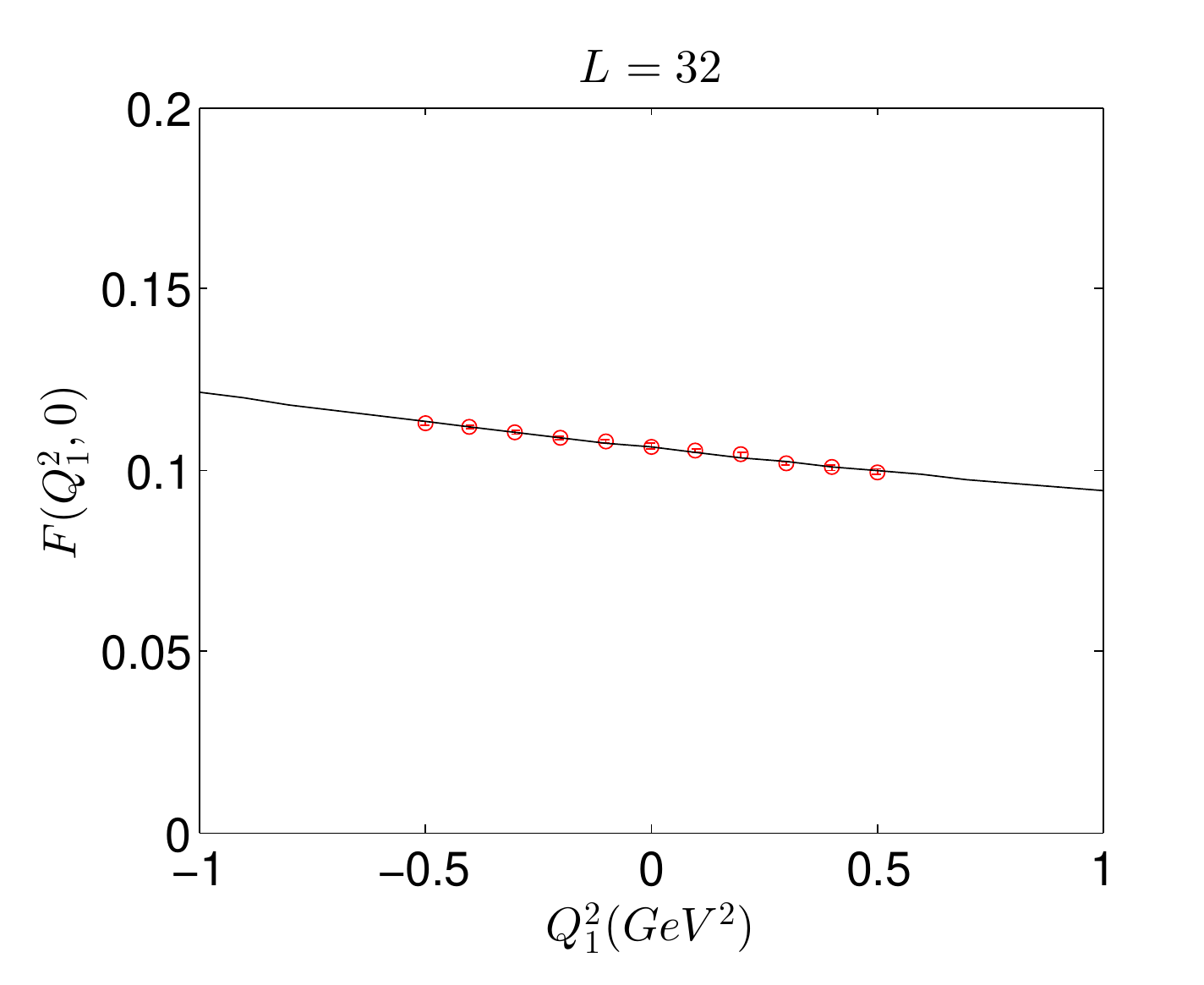}}
  \centerline{Ensemble II}
\end{minipage}
\caption{$F(Q_1^2, 0)$ is again fitted with a one-pole form:
 $F(Q_1^2, 0)=F(0,0)/(1+Q_1^2/\nu^2)$ for Ensemble I (left figure)
 and Ensemble II (right figure). }
\label{fig:FormFactor_fit2}
\end{figure*}
 Having obtained the results for $F(Q_1^2, 0)$, we can fit it again with another one-pole form,
 \be
 F(Q_1^2, 0)=F(0,0)/(1+Q^2_1/\nu^2)
 \label{eq:formfactor_fit_2}
 \ee
 with $F(0,0)$ and $\nu^2$ being the fitting parameters.
 This is illustrated in Fig.~\ref{fig:FormFactor_fit2} for two of our ensembles.
 Again, correlated fits are adopted here.

\begin{figure*}
  \hspace{-10mm}
  \includegraphics[width=9.5cm]{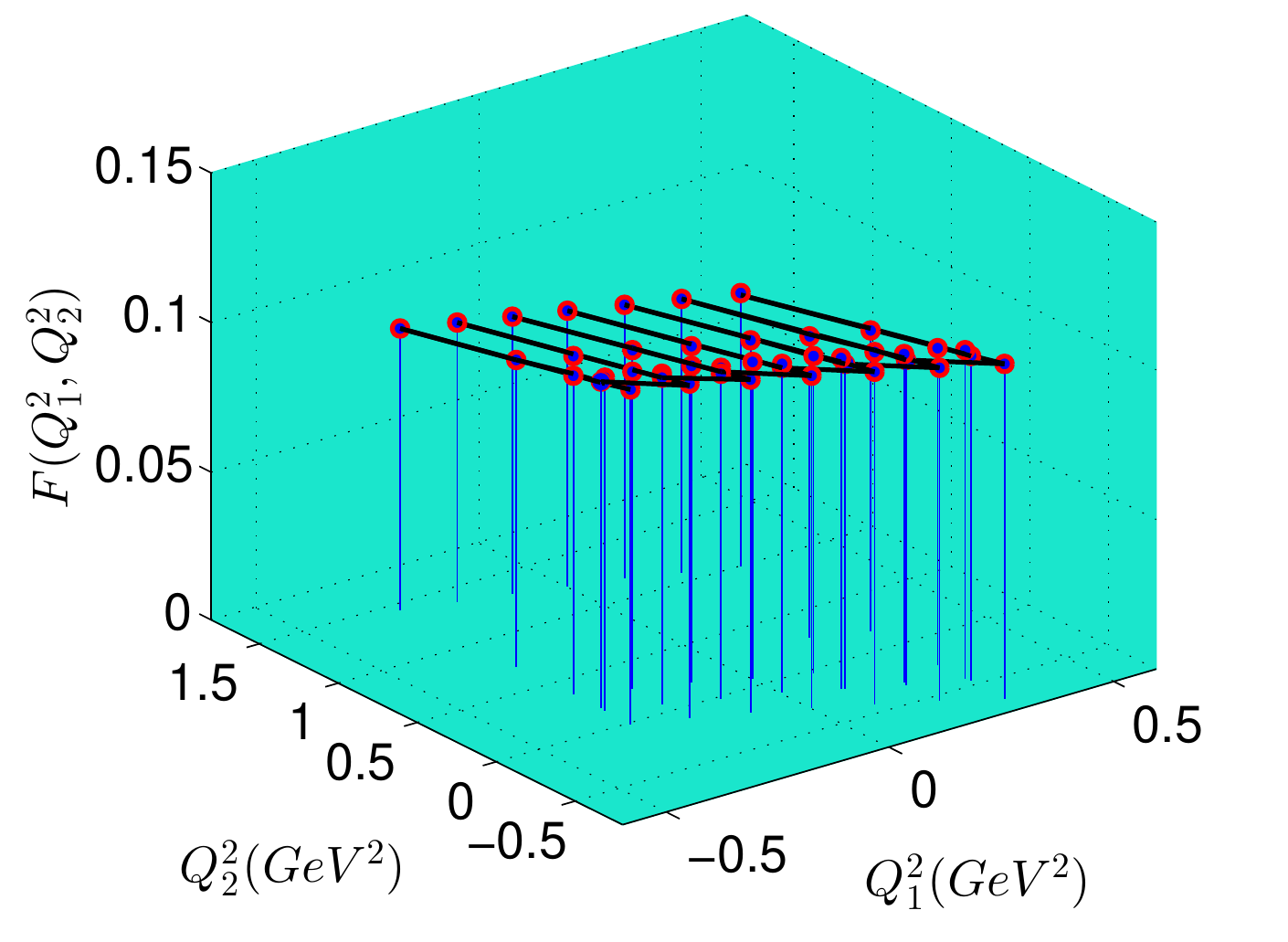}
  \hspace{-10mm}
  \includegraphics[width=9.5cm]{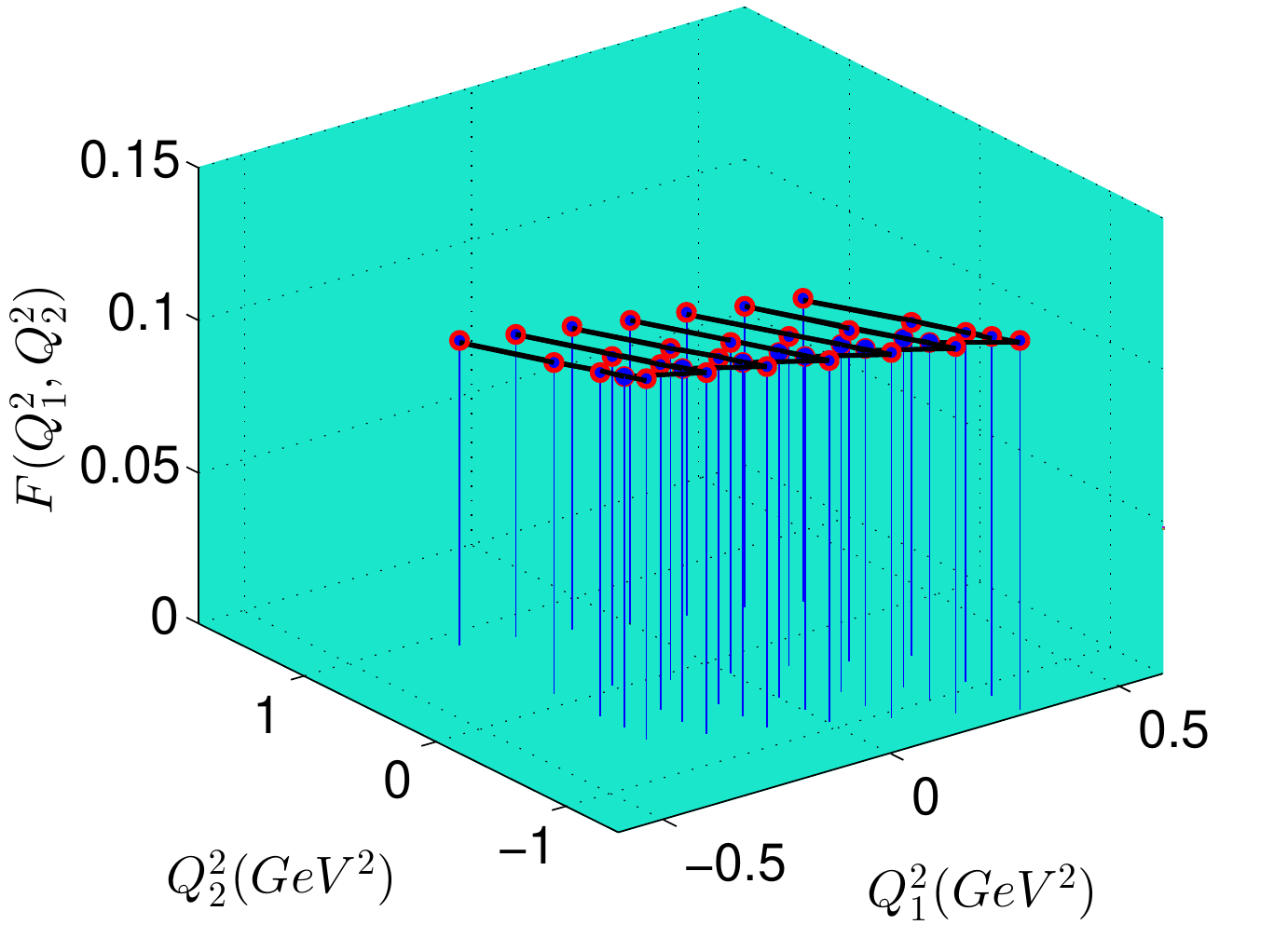}
\caption{$F(Q_1^2, Q_2^2)$ is fitted with Eq.~(\ref{eq:formfactor_fit_3}) are shown for  Ensemble I (left figure)
 and Ensemble II (right figure). }
\label{fig:FormFactor_fit3}
\end{figure*}

 Apart from fitting the data in a two-step procedure as described above,
 we have also tried to fit the data in a one-step method.
 When we plug Eq.~\ref{eq:formfactor_fit_2} into Eq.~\ref{eq:formfactor_fit_1} and
 assuming that we are only interested in the value of the form factor close
 to the physical point, we may Taylor expand it assuming both $Q_1^2$ and $Q_2^2$ are small,
 \beq
  \label{eq:formfactor_fit_3}
F(Q_1^2,Q_2^2)=F(0,0)+aQ_1^2+bQ_2^2\;,
\;\;\;Q^2_1,Q^2_2\sim 0\;.
\eeq
 Thus, we could fit the data in a region close to the origin
 with $F(0,0)$, $a$ and $b$ being the fitting parameters.
 This is illustrated in Fig.~\ref{fig:FormFactor_fit3} for two of our ensembles.
 In each case, $35$ data points of $(Q^2_1,Q^2_2)$ close to the origin are
 taken and the corresponding form factors $F(Q^2_1,Q^2_2)$ are obtained.
 Then using a linear fit in both
 $Q^2_1$ and $Q^2_2$, c.f. Eq.~(\ref{eq:formfactor_fit_3}), the form factors
 at the origin are obtained for both ensembles. Again, correlated fits are adopted here.
 The fitting results are summarized in Table~\ref{final_fit_1}.

 When computing the physical double photon decay width, according to Eq.~(\ref{decay_width}),
 one has to plug in the mass of the $\eta_c$ meson. What we really compute on
 the lattice is the combination of correlation functions which is
 related to the matrix element $\langle \eta_c| \gamma(q_1,\lambda_1)\gamma(q_2, \lambda_2)\rangle$
 via Eq.~(\ref{master}). When we parameterize this particular matrix element
 in terms of form factor in Eq.~(\ref{eq:formfactor_parametrization}),
 the relation involves $m_{\eta_c}$ as well. Therefore, the decay width
 turns out to be proportional to $m^3_{\eta_c}$:
 $\Gamma\propto m^3_{\eta_c}|\langle \eta_c| \gamma(q_1,\lambda_1)\gamma(q_2, \lambda_2)\rangle|^2$.
 Here it is then quite different if one substitutes in the value of $m_{\eta_c}$ obtained on the lattice,
 or the true physical value of $m^{\rm phys.}_{\eta_c}=2.98$GeV,
 the two differs by about 10\% for the coarser lattice
 and about 5\% for the finer lattice.
 Therefore, if one would substitute in the true physical mass,
  it will result in a 15\% difference in the value of $\Gamma$ for the finer lattice
  and about 30\% for the coarser one.

 The reason for the above mentioned difference is the following.
 We are taking the value of the valence charm quark mass parameter $\mu_c$ from
 Ref.~\cite{Becirevic:2012dc}. There, it is assumed that, when the continuum limit is taken,
 the value of $m_{\eta_c}$ will recover its physical value. However, being on a
 finite lattice, the computed value of $m_{\eta_c}$ comes out to be less than
 the corresponding physical value. The difference of the two is in fact an estimate
 of the finite lattice spacing error. In fact, $m_{\eta_c}$ is not the
 only factor which affects the results. The renormalization factor $Z_V(g^2_0)$ that
 we quoted in Table~\ref{configurations} also depends on the lattice spacing.
 Therefore, we think it is more consistent to substitute in the values of $m_{\eta_c}$
 computed on each ensembles.
 In the end, of course, one should try to take the continuum limit when
 the lattice computations are performed on a set of ensembles with different lattice spacings.

 If using the two-step fitting procedure using Eq.~(\ref{eq:formfactor_fit_1}) and using Eq.~(\ref{eq:formfactor_fit_2}), with the values of $m_{\eta_c}$ obtained from each ensemble substituted in,
 we obtain for the decay width $\Gamma=1.019(3)$KeV for the coarser
 and  $\Gamma=1.043(3)$KeV for the finer lattice ensembles.
 These results for the form factor $F(0,0)$ together with
 the corresponding results for the decay width are summarized in Table~\ref{final_fit}.

 As for the one-step fitting procedure using Eq.~(\ref{eq:formfactor_fit_3}), we obtain
 $\Gamma=1.025(5)$KeV for the coarser and  $\Gamma=1.062(5)$KeV for the finer lattice ensembles.
 The results for the form factor $F(0,0)$ and $\Gamma(\eta_c\rightarrow\gamma\gamma)$ are
 consistent with each other for Ensemble I using two different types of fitting procedure.
 However, for Ensemble II, a combined fitting using Eq.~(\ref{eq:formfactor_fit_3}) gives
 a larger result for both $F(0,0)$ and $\Gamma(\eta_c\rightarrow\gamma\gamma)$.
 We think the value using the combined fit is more reliable since it gives a much
 less value of $\chi^2/dof$. In the combined fitting, the naive continuum extrapolated
 result for the decay width reads $\Gamma=1.122(14)$KeV.
 The fitted results for $F(0,0)$ together with the corresponding results
 for the decay width  are summarized in Table~\ref{final_fit_1}.

 Let us now discuss the possible systematic errors.
 Although the mass of the pion in the two ensembles are relatively heavy,
 we do not expect the double photon decay width to be very sensitive
 to the pion mass. Also, since both of our ensembles have $m_\pi L\sim 3.3$,
 we do not expect very large finite volume errors as well.
 Since we have only two ensembles, it is not possible to
 make reliable extrapolation towards the continuum limit.
 However, if one would try a naive continuum limit extrapolation,
 assuming an $\calO(a^2)$ error, we obtain $\Gamma=1.082(10)$KeV
 which is also listed in Table~\ref{final_fit}. There are of course
 other sources of systematic errors, e.g. the neglecting of the so-called disconnected
 contributions, the quenching of the strange quark, etc.
 Therefore, we decided not to quantify the systematic errors in this exploratory study.
 However, as we discussed above, the difference in the $\eta_c$ mass
 already indicates that there might be a finite lattice spacing error
 at the order of 15\% for the finer and 30\% for the coarser ensembles,
 respectively.

\begin{table*}
\begin{center}
\caption{$F(Q_1^2, 0)$ is again fitted with a one-pole form:
 $F(Q_1^2, 0)=F(0,0)/(1+Q_1^2/{\nu^2})$ are shown for the two ensembles
 (the first two lines).
 In the last column, we show the decay width obtained using Eq.~(\ref{decay_width}).
 A naive continuum extrapolation are shown in the third line for reference.}
 \label{final_fit}
\begin{tabular}{|c|c|c|c|c|}
\hline
 &$F(0,0)$     &$\nu^2(GeV^2)$    &$\chi^2/dof$  &$\Gamma(\eta_c \to \gamma\gamma)$(KeV)\\
 \hline
Ensemble I    &0.10719(13)       &7.13(20) &0.19/9    &1.019(3)\\
\hline
Ensemble II   &0.10608(17)       &7.88(29) &3.4/9     &1.043(3)\\
\hline
Naive extrapolation          &             &                   &              &1.082(10)\\
\hline
\end{tabular}
\end{center}
\end{table*}

\begin{table*}
\begin{center}
\caption{$F(Q_1^2, Q_2^2)$ is fitted with Eq.~(\ref{eq:formfactor_fit_3}) are shown for the two ensembles
 (the first two lines).
 In the last column, we show the decay width obtained using Eq.~(\ref{decay_width}).
 A naive continuum extrapolation are shown in the third line for reference.}
 \label{final_fit_1}
\begin{tabular}{|c|c|c|c|c|c|}
\hline
 &$F(0,0)$     &$a$    &$b$ &$\chi^2/dof$  &$\Gamma(\eta_c \to \gamma\gamma)$(KeV)\\
 \hline
Ensemble I    &0.10750(24)     &-0.0138(11)     &-0.01216(36)    &2.67/32
&1.025(5)\\
\hline
Ensemble II   &0.10705(24)     &-0.0124(11)     &-0.01282(38)    &2.50/32
&1.062(5)\\
\hline
Naive extrapolation          &             &                  &  &
&1.122(14)\\
\hline
\end{tabular}
\end{center}
\end{table*}

 Now let us briefly discuss the implications of our lattice results for $\Gamma$.
 First of all, our results are substantially smaller than
 the previously obtained quenched lattice result in Ref.~\cite{Dudek:2006ut},
 which we quote: $\Gamma=2.65(26)_{stat}(80)_{scal.}(53)_{quen.}$KeV.
 Second, our value is also much smaller than most of the experimental values.
 The current experimental value, according to PDG, is about $5.0$~KeV with
 an error of $0.4$~KeV~\cite{Agashe:2014kda}.
 However, one should note that the most recent determination of this quantity
 by Belle~\cite{Zhang:2012tj} with the
 result $5.8\pm 1.1$~KeV is not a direct measurement of $\Gamma_{\gamma\gamma}$ itself,
 but the product $\Gamma_{\gamma\gamma}B(\eta_c\rightarrow\eta'\pi^+\pi^-)\simeq 50.5$~eV.
 Therefore, the value of $\Gamma_{\gamma\gamma}$ is usually extracted
 by inferring to earlier measurements in other channels, making the final
 results for $\Gamma_{\gamma\gamma}$ differ quite a bit. For example, if we blindly use
 the PDG quoted value of the branching ratio, $B(\eta_c\rightarrow\eta'\pi^+\pi^-)=0.041\pm 0.017$,
 we arrive at $\Gamma_{\gamma\gamma}\simeq 1.25$~KeV, which is comparable to our lattice result.
 However, if we would infer
 $\Gamma_{\gamma\gamma}$ from the ratio of
 $\Gamma_{\gamma\gamma}\Gamma(K\bar{K}\pi)/\Gamma_{tot}=0.407\pm 0.027$KeV and
 $\Gamma(K\bar{K}\pi)/\Gamma_{tot}=(7.0\pm1.2)\times 10^{-2}$, we end up with
 $\Gamma_{\gamma\gamma}=5.8\pm1.1$KeV as in Ref.~\cite{Zhang:2012tj}. Therefore, it is highly desirable to
 have a more  precise and/or direct measurement of this quantity in future experiments.

 The possible reasons for these apparent discrepancies can come from several sources
 to be discussed below. First, we have used different configurations from the quenched calculations.
 Our calculation takes into account the sea quark contributions from $u$ and $d$ quarks while
 in Ref.~\cite{Dudek:2006ut} these have been ignored. Although the quenching errors have
 been estimated in Ref.~\cite{Dudek:2006ut}, it is well-known that this type of systematic
 is very difficult to quantify accurately. It is therefore quite possible that
 these effects have been under-estimated in Ref.~\cite{Dudek:2006ut}.

 Another possibility is that we have a rather large systematic errors
 which is not fully quantified in this exploratory study. It is seen that our statistical
 errors seem to be small. However, as mentioned above, we do observe a large finite lattice spacing error of
 about 15-30\% just from the mass of the $\eta_c$. Since we have only two lattice spacings,
 the continuum limit extrapolation is also not well-controlled.
 In fact, if we blindly ascribe an error of about 15\% for the numbers
 of the decay width for the two ensembles in Table~\ref{final_fit},
 it is possible that we could end up with a number that is
 close to the quenched result but with a rather large error
 coming from the continuum limit extrapolation. Of course, it is also possible
 that this disagreement is due to the combination of the above mentioned sources.
 In any case, a more systematic study with more lattice ensembles will
 definitely help to clarify these issues.

\section{Conclusions}
 \label{sec:conclude}

 In this exploratory study, we calculate the decay width for two-photon decay of $\eta_c$ using
 unquenched $N_f=2$ twisted mass fermion configurations.
 The computation is done with two lattice ensembles
 at two different lattice spacings.
 The mass spectrum and dispersion relations for the $\eta_c$ state are first examined.
 It is verified that lattice dispersion relations are
 better than the continuum ones. The implication of this is carried over
 to the computation of hadronic matrix element and the corresponding form factors.

 By calculating various three-point functions, two-photon decays of $\eta_c$ matrix element
 are obtained at various of virtualities. It is particularly helpful to implement
 the so-called twisted boundary conditions which enable us to populate the physical region well.
 The matrix element is decomposed into kinematic factors and one form factor $F(Q_1^2,Q_2^2)$
 which is obtained in a region close to the physical point.
 Then, we adopt a simple one-pole parametrization to fit the data for each value of $Q_1^2$,
 and subsequently fit $F(Q_1^2, 0)$  again with a one-pole form yielding the value of $F(0,0)$.
 A naive continuum extrapolation gives $\Gamma=1.082(10)$KeV.
 We also use the Taylor expansion for $F(Q_1^2,Q_2^2)$ with respect to $Q_1^2$ and $Q_2^2$
 close to the origin and extract the value of $F(0,0)$,
 the naive continuum extrapolation of which yields $\Gamma=1.122(14)$KeV.

 Our result is significantly smaller than both the quenched result and the experimental values
 quoted by the PDG. However, taking into account of the possibly large systematic errors
 in the present lattice computations and the large uncertainties in the experimental result
 itself, it is still premature to say that there is a severe discrepancy here. Obviously,
 future more systematic lattice studies with various lattice spacings
 and more statistics are very much welcome here. It would also be helpful to
 estimate the disconnected contributions that has been neglected in this exploratory study.
 It will also be helpful to use other types of unquenched configurations, e.g. with $2+1$ flavors
 or even $2+1+1$ flavors in order to estimate the effects for the quenching of the other quark flavors.
 Last but not the least, more precise experimental results on double photon decays of charmonium
 are crucial in this area as well.

 \section*{Acknowledgments}

 The authors would like to thank the European Twisted Mass Collaboration (ETMC)
 to allow us to use their gauge field configurations. Our thanks also go to
 National Supercomputing Center  in Tianjin (NSCC) and the Bejing Computing Center (BCC) where part of the numerical computations are performed.
 This work is supported in part by the
 National Science Foundation of China (NSFC) under the project
 No.11505132, 
 No.11335001, 
 No.11275169, 
 No.11405178, 
 No.11575197. 
 It is also supported in part by the DFG and the NSFC (No.11261130311) through funds
 provided to the Sino-Germen CRC 110 ``Symmetries and the Emergence
 of Structure in QCD''.
 This work is also funded in part by National Basic Research Program of China (973 Program)
 under code number 2015CB856700.
 M.~Gong and Z.~Liu are partially supported by the
 Youth Innovation Promotion Association of CAS (2013013, 2011013).
 This work is also supported by the Scientific Research Program Funded by Shaanxi
Provincial Education Department under the grant No.~15JK1348, and Natural Science Basic Research Plan in
Shaanxi Province of China (Program No.~2016JQ1009).

%


%

 \end{document}